\documentstyle[12pt,a4,amstex,graphics]{article}

\newcommand{\ga}{\alpha}
\newcommand{\gb}{\beta}
\renewcommand{\gg}{\gamma}
\newcommand{\gd}{\delta}
\renewcommand{\ge}{\epsilon}
\newcommand{\gve}{\varepsilon}

\newcommand{\gx}{\xi}
\newcommand{\gm}{\mu} 
\newcommand{\gn}{\nu}

\newcommand{\gk}{\kappa}
\newcommand{\gl}{\lambda}
\newcommand{\gr}{\rho}
\newcommand{\gth}{\theta}
\newcommand{\gs}{\sigma}

\newcommand{\go}{\omega}

\newcommand{\gps}{\psi}
\newcommand{\get}{\eta}
\newcommand{\gch}{\chi}
\newcommand{\gG}{\Gamma}
\newcommand{\gD}{\Delta}
\newcommand{\gF}{\Phi}

\newcommand{\gL}{\Lambda}
\newcommand{\gS}{\Sigma}

\newcommand{\gO}{\Omega}

\newcommand{\gPs}{\Psi}

\newcommand{\cA}{{\cal A}}
\newcommand{\cB}{{\cal B}}
\newcommand{\cC}{{\cal C}}
\newcommand{\cD}{{\cal D}}
\newcommand{\cE}{{\cal E}}
\newcommand{\cF}{{\cal F}}
\newcommand{\cG}{{\cal G}}

\newcommand{\cK}{{\cal K}}
\newcommand{\cL}{{\cal L}}
\newcommand{\cM}{{\cal M}}

\newcommand{\cR}{{\cal R}}

\newcommand{\cW}{{\cal W}}

\newcommand{\uA}{{\underline A}}
\newcommand{\uB}{{\underline B}}
\newcommand{\uC}{{\underline C}}
\newcommand{\uD}{{\underline D}}

\newcommand{\tg}{{\tilde g}}

\newcommand{\tH}{{\tilde H}}

\newcommand{\tP}{{\tilde P}}

\newcommand{\uga}{{\underline\alpha}}
\newcommand{\ugb}{{\underline\beta}}
\newcommand{\ugg}{{\underline\gamma}}

\newcommand{\ucA}{{\underline{\cal A}}}
\newcommand{\ucB}{{\underline{\cal B}}}
\newcommand{\ucC}{{\underline{\cal C}}}
\newcommand{\ucD}{{\underline{\cal D}}}
\newcommand{\ucE}{{\underline{\cal E}}}

\newcommand{\bs}{{\bar s}}

\newcommand{\bx}{{\bar x}}

\newcommand{\bz}{{\bar z}}
\newcommand{\bA}{{\bar A}}

\newcommand{\bD}{{\bar D}}
\newcommand{\bE}{{\bar E}}
\newcommand{\bF}{{\bar F}}

\newcommand{\bH}{{\bar H}}

\newcommand{\bL}{{\bar L}}

\newcommand{\bQ}{{\bar Q}}
\newcommand{\bR}{{\bar R}}

\newcommand{\bU}{{\bar U}}
\newcommand{\bV}{{\bar V}}

\newcommand{\bX}{{\bar X}}

\newcommand{\bZ}{{\bar Z}}
\newcommand{\bga}{{\bar \alpha}}

\newcommand{\bge}{{\bar\epsilon}}

\newcommand{\bgl}{{\bar\lambda}}

\newcommand{\bgps}{{\bar\psi}}

\newcommand{\bgch}{{\bar\chi}}
\newcommand{\bgG}{{\bar\Gamma}}

\newcommand{\bgF}{{\bar\Phi}}

\newcommand{\bgS}{{\bar\Sigma}}

\newcommand{\bgO}{{\bar\Omega}}

\newcommand{\bgPs}{{\bar\Psi}}

\newcommand{\bcF}{{\bar{\cal F}}}

\newcommand{\bcR}{{\bar{\cal R}}}

\newcommand{\tr}{\text{tr}}

\newcommand{\slashed}{\hspace{-1.1ex}/}
\newcommand{\Slashed}{\hspace{-1.3ex}/\hspace{.2ex}}
\newcommand{\lra}{\longrightarrow}

\renewcommand{\Re}{\mbox{Re}}
\renewcommand{\Im}{\mbox{Im}}

\newcommand{\der}{\partial}
\newcommand{\Der}{D}
\newcommand{\sDer}{\Der\Slashed}
\newcommand{\sder}{\der\slashed}
\newcommand{\lrset}[1]{{\overset{\leftrightarrow}{#1}}}
\newcommand{\inv}{^{-1}}
\newcommand{\nit}{\noindent}
\newcommand{\nl}{\newline}
\newcommand{\np}{\newpage}
\newcommand{\dsp}{\displaystyle}

\newcommand{\ct}{\cite}

\newcommand{\lh}{\left(}
\newcommand{\rh}{\right)}
\newcommand{\ubar}[1]{{\underline{#1}}}

\newcommand{\labl}[1]{\label{#1}}
\newcommand{\half}{\frac 12 }

\newcommand{\Kh}{K\"{a}hler} 
\newcommand{\beq}{\begin{equation}}
\newcommand{\eeq}{\end{equation}}
\newcommand{\barr}{\begin{array}}
\newcommand{\earr}{\end{array}}
\newcommand{\equ}[1]{\begin{gather} #1 \end{gather}}
\newcommand{\equa}[1]{\begin{align} #1 \end{align}}

\newcommand{\pmtrx}[1]{\begin{pmatrix} #1 \end{pmatrix}}

\newcommand{\non}{\nonumber}
\newcounter{oldcounter}

\numberwithin{equation}{section}
\begin{document}

\pagestyle{empty}

\begin{flushright}
NIKHEF/99-007 
\end{flushright} 

\begin{center}
{\Large 
{\bf Consistent $\boldsymbol{\sigma}$-models in N=1 supergravity}} \\
\vspace{5ex}

{\large S.\ Groot Nibbelink and J.W.\ van Holten}\\
\vspace{3ex}

{\large NIKHEF, P.O.\ Box 41882,}\\
\vspace{3ex}

{\large 1009 DB, Amsterdam NL}
\vspace{5ex}

{{\em Revised version} \\
 August 8, 1999}
\end{center}
\vspace{15ex}

\nit
{\small 
{\bf Abstract}
\nl
A consistent $N=1$ supersymmetric $\sigma$-model can be constructed, 
given a K\"ahler manifold, by adding chiral matter multiplets. Their 
scalar components are covariant tensors on the underlying K\"ahler 
manifold. The K\"ahler $U(1)$-charges can be adjusted such 
that the anomalies cancel, using the holomorphic functions in 
which the K\"ahler potential transforms. The arbitrariness of 
the $U(1)$-charges of matter multiplets is related to their 
Weyl-weights in superconformal gravity, before it is reduced to 
supergravity. The covariance of the K\"ahler potential forces the 
superpotential to be covariant as well. This relates the cut-off, 
the Planck scale and the matter charges to each other. 
A non-vanishing VEV of the covariant superpotential 
breaks the K\"ahler $U(1)$ spontaneously. If this VEV vanishes, 
the gravitino is massless and depending on the above mentioned 
parameters there may be additional internal symmetry breaking. 
The separation of the different representations of chiral 
multiplets can be achieved by covariantizations of derivatives 
and fermions. Using non-holomorphic transformations, the full 
K\"ahler metric can be block-diagonalized and the necessary 
covariantizations come out naturally. Various aspects are 
illustrated by applying them to Grassmannian coset models. 
As an example the coset $SU(5)/SU(2)\times U(1) \times SU(3)$ 
with the field content of the standard model is constructed. 
Phenomenological aspects of this model are analyzed.
} 

\np

\pagestyle{plain}
\pagenumbering{arabic}

\section{Introduction}
\label{Intro}

Like gravity, non-linear $\gs$-models in four dimensions are not 
renormalizable. They involve a natural scale $\gL$ above which 
new dynamics or new physical degrees of freedom can come into play. 
The existence of such a scale requires the introduction of a 
parameter $f \propto \gL\inv$ of inverse mass dimension. Below 
this scale the $\gs$-models are useful as effective field theories, 
for example to describe the low-energy dynamics of bound states 
in strongly interacting gauge theories like QCD. Effective 
field theories also arise in the low-energy description of quantum
string theory, in which case the string scale sets the limit of 
applicability. This scale is connected with the Planck scale, as 
classical gravity described in a general relativistic formulation 
is part of the effective long-distance physics which comes out of 
string theory. 

Because realistic string theories require supersymmetry for their 
consistency, the most obvious candidates for effective low-energy 
theories arising from string models are four-dimensional $N = 1$ 
supergravity theories, subject to phenomenological constraints 
as well. These theories describe gravity and the other 
interactions in the context of a locally supersymmetric field theory, 
which is not renormalizable but well-behaved below the Planck scale. 
With well-behaved we mean, that by taking into account the 
presence of a cut-off, these theories give unambiguous and 
consistent answers to questions related to phenomena at distance 
scales large compared to the cut-off. In particular one requires
the proper incorporation of symmetries and the absence of anomalies 
in local gauge-invariances like those of the electro-weak or 
grand-unified interactions. 

For such reasons it is important to be able to construct the 
most general locally supersymmetric field theories including 
local chiral and non-chiral gauge-interactions free of anomalies 
at the quantum level. In this paper we address this question 
within the context of conventional representations of $N = 1$ 
supersymmetry, which besides the supergravity multiplet include 
complex chiral and real vector multiplets. 

The complex scalars of the chiral-multiplet sector of an $N = 1$ 
supersymmetric theory parameterize a \Kh\ manifold.
In particular, the low-energy lagrangean of this sector is 
determined by a single real function of the chiral superfields 
$\gS^{{\cal A}}$, the \Kh\ potential $\cK(\bgS, \gS)$, from 
which objects like the metric and curvature of the manifold 
can be computed. Interesting examples of \Kh\ manifolds include 
cosets of the Grassmannian type \cite{Ong:1983te, Mattis:1983wn, 
vanHolten:1985di, Achiman:1984ku}, 
and those based on exceptional groups like $E_6/SO(10) 
\times U(1)$, $E_7/SU(5) \times SU(3) \times U(1)$ or 
$E_8/SO(10)\times SU(3) \times U(1)$ 
\cite{Achiman:1985ra, Bando:1988br, Yanagida:1985jc, Kugo:1984ai, 
Ong:1983uj}. 

The pure $N = 1$ supersymmetric $\gs$-models can be extended in 
several ways: by adding superpotentials, by gauging some or all 
of the isometries and by adding additional matter superfields in 
appropriate representations of the isometry group. For 
model-building purposes it is necessary to analyze what kind 
of low-energy physics then emerges from these models and their 
various extensions. This is a highly non-trivial question, as 
supersymmetry requires the inclusion of many rather special 
scalar potential terms and Yukawa couplings for its consistency. 
The patterns of internal and supersymmetry breaking, and of 
boson and fermion masses emerging from these models can become 
quite intricate.

Further constraints come from requiring the light particles in 
the model to have assigned charges and other quantum 
numbers to them, consistent with standard-model phenomenology, 
and to be free of chiral anomalies in the $\gs$-model or gauge 
interactions. This has been the subject of various earlier 
investigations \ct{Moore:1984dc,Cohen:1985js, Shore:1989mn, 
Kotcheff:1990ck, Buchmuller:1987zp}. 
In these studies it was concluded that many models based on 
mathematically interesting coset spaces like the symmetric 
spaces, appear to suffer from problems in these respects. 
However, in a recent paper \cite{GrootNibbelink:1998tz} we 
showed that the $U(1)$ charge assignments of 
the chiral superfields allow more freedom then was previously 
thought. As a result anomaly cancellation can be realized 
in phenomenologically interesting ways by combining 
appropriate representations of isometry groups, even in models 
based on these coset spaces. It is then of clear importance to 
study which consistent effective supergravity models can be 
constructed incorporating these ideas. 

The present paper is a step in this direction. Building on the 
results of \ct{Cremmer:1983en, Kugo:1983mr, Bagger:1983ab}, 
we discuss the extension of the supergravity lagrangeans 
necessary to include non-linear internal gauge symmetries. 
We describe how the non-linear transformations can be modified 
to assign arbitrary $U(1)$ charges to supermultiplets in any 
representation of the internal gauge group, and we discuss the 
cancellation of anomalies. 

This paper is composed as follows. The main aspects of gauged 
supersymmetric $\gs$-models on \Kh\ manifolds are reviewed in 
section \ref{O}. We discuss their extension to local supersymmetry 
in section \ref{VII}. This includes a description of the role of 
the non-linear compensating scalar multiplet introduced in 
\cite{GrootNibbelink:1998tz}. 
Section \ref{XI} analyses the phenomenology in supergravity of 
models where the \Kh\ potential transforms covariantly. 
In section \ref{VI} the various constructions are applied in 
the context of the Grassmannian coset spaces $U(n+m)/U(n) 
\times U(m)$ and its non-compact analogs. 
An anomaly-free supersymmetric model of a family of quarks and 
leptons in representations of $SU(5)/SU(3) \times SU(2) \times 
U(1)$ is presented in sect.\ref{VI.sub1}. Diagonalization of 
the metric and propagators is discussed in a general geometrical 
setting in section \ref{II}, using some of the geometrical 
constructions presented in the appendix. Its implementation 
in the case of Grassmannian models is given in section \ref{VIII}. 
The vacuum configuration of the Grassmannian model based on 
$SU(5)$ with the standard model particle spectrum is analysed.
The conclusions and lessons drawn from these investigations 
are summarized in section \ref{X}. Their applications to other 
models are described in a separate publication \cite{sjwinpre}. 

\section{Supersymmetric lagrangeans}
\label{O}

In this section we present the machinery to describe $N=1$ 
supersymmetric lagrangeans \ct{Bagger:1982, Ong:1983te, 
Mattis:1983wn, vanHolten:1985di}. The geometrical objects we 
use in this section are just short-hand to cast the formulae in
a more systematic form. The geometrical aspects are of use 
later on in this article. All the supersymmetric field theories 
which are developed in this section have to be interpreted as 
effective field theories involving a cut-off scale $f\inv$. 
This cut-off scale is used 
explicitly only when needed to give a certain object its 
canonical dimension. 

Let $\gS^\cA = (Z^\cA, \gps_L^\cA, H^\cA)$ be a set of chiral 
multiplets, where $Z^\cA$ is a physical complex scalar, 
$\gps_L^\cA$ a chiral fermion and $H^\cA$ is an auxiliary complex 
scalar. The index $\cA$ enumerates the multiplets in the set.  
The kinetic part of the lagrangean for such chiral multiplets 
is given in terms of a real composite superfield $\cK(\bgS, \gS)$ 
by the following supersymmetric expression 
\equ{
\cL_\cK \, =\, \cK(\bgS, \gS)|_D\, =\, 
 - g_{\ucA \cA}\, \lh \der^{\gm} \bZ^{\ucA} \der_{\gm} 
 Z^{\cA} + \bgps_L^{\ucA}\, \lrset\Der\Slashed \gps_L^{\cA} -
\hat {\bH}^\ucA \hat H^\cA_{\;}
\rh
\non \\
\labl{O.2.2}\labl{O.1}  \\
 +\, R_{\ucA \cA\ucB \cB}\, 
(\bgps_R^{\cA} \gps_L^{\cB}) \; (\bgps_L^{\ucA} \gps_R^{\ucB}).
\non
}  
The complex hermitean metric $g_{\ucA\cA}$ can be derived from 
the \Kh\ potential $\cK(\bZ,Z)$ by
\begin{equation}
g_{\ucA\cA}\, =\, \cK_{, \ucA \cA}.
\labl{O.2}
\end{equation} 
The auxiliary fields $H^\cA$ are redefined as
\(
\hat H^\cA = H^\cA - \gG^\cA_{\cB\cC} \bgps^\cB_R \gps^\cC_L
\)
and the \Kh\ covariant derivative 
\(
\Der_\gm \gps_L^\cA = 
\der_\gm \gps_L^\cA + \gG^\cA_{\cB\cC} \gps_L^\cB \der_\gm Z^\cC
\)
is introduced with 
\equ{
\gG^\cA_{\cB\cC} = g^{\cA \ucA} g_{\ucA\cB, \cC}, \quad
\bgG^\ucA_{\ucB \, \ucC} = g^{\cA \ucA} g_{\ucB\cA, \ucC}
\labl{O.2.0.2}
}
\nit
the complex connections. The four-fermion terms have been 
rewritten with the help of the curvature tensor 
\equ{
R_{\ucA\cA\ucB\cB} = g_{\ucA\cA,\ucB\cB} - g_{\ucA\cC,\cA} \, 
g^{\cC\ucC} \, g_{\ucC\cB,\ucB}.
\labl{O.2.1}
} 
In addition one can write down a lagrangean determined by a 
holomorphic function $W$, called the superpotential, 
of chiral superfields
\equ{
\cL_{W} \,=\, \left[ W(\gS) \right]_F = \half W_{,\cA} H^\cA 
- \half W_{, \cA\cB} \bgps_R^\cA \gps_L^\cB + \mbox{h.c.}.
\labl{O.3} 
} 
\nit
It follows from the lagrangean \eqref{O.1} that the symmetries of 
this supersymmetric model are given by the isometries of the metric 
$g_{\ucA\cA}$ which leave \eqref{O.3} invariant\footnote{In fact 
$W$ may transform with a  phase factor which does not depend on 
the fields, which can be compensated by a chiral rotation 
of the fermions to leave the Yukawa couplings invariant. This 
so-called $R$-symmetry is broken if the superpotential does not 
transform homogeneously.}. 
The isometries are generated by a complete set of the Killing 
vectors $\cR_i^\cA(Z)$ which determine the transformation rules 
for the chiral multiplet completely
\equ{
\gd_i \gS^\cA = \cR_i^\cA(\gS) = 
\left\{
\barr{ll}
\gd_i Z^\cA & = \cR_i^\cA(Z), \\
&\\
\gd_i \gps_L^\cA & =  \cR^\cA_{i,\cB}(Z) \gps_L^\cB, \\
& \\
\gd_i H^\cA & = \cR^{\cA}_{i,\cB}(Z) H^\cB - \cR^\cA_{i,\cB\cC}(Z)
\bgps_R^\cB \gps_L^\cC.
\earr
\right.
\labl{O.5.1}\labl{O.5.2}\labl{O.4}
} 
The Killing vectors satisfy the Killing conditions
\(
\lh g_{\ucA \cB} \cR_i^\cB \rh_{,\cA} + 
\lh \bcR_i^\ucB g_{\ucB \cA}\rh_{,\ucA} = 0,
\)
therefore they obey 
\equ{
\cR^\cB_{[i} \cR^\cA_{j],\cB} = f_{ij}^{\;\;\;k} \cR^\cA_k.
\labl{O.6}
} 
This defines a representation of the abstract algebra 
\(
\gd_{[i} \gd_{j]} = f_{ij}^{\;\;\;k} \gd_k
\)
of a group with the structure coefficients $f_{ij}^{\;\;\;k}$, 
satisfying the Jacobi identities for consistency.

The K\"ahler potential $\cK$ may transform under the isometries 
\eqref{O.4} as
\equ{
\gd_i \cK(\bZ, Z) =  \cK_{,\cA} \cR_i^\cA + \cK_{,\ucA} 
\bcR_i^\ucA =  \cF_i(Z) + \bcF_i(\bZ).
\labl{O.7}
} 
The functions $\cF_i$ ($\bcF_i$) are (anti-)holomorphic functions, 
as the metric is defined by \eqref{O.2}.
By using the group property of the isometries and the fact that 
$\cF_i$ and $\cR_i$ are both holomorphic, it follows that the 
algebra of the functions $\cF_i$ is determined by the structure 
constants, up to an imaginary constant part: 
\(
\gd_{[i} \cF_{j]} = f_{ij}^{\;\;\;k} \cF_k + i a_{ij}
\)
where the constants $a_{ij}$ are real and anti-symmetric. 
By an appropriate shift of the functions $\cF_i$ these constants 
can be absorbed into the definition of $\cF_i$, so as to give 
\equ{
\gd_{[i} \cF_{j]} = f_{ij}^{\;\;\;k} \cF_k.
\labl{O.7.1} 
} 
Thus the holomorphic functions $\cF_i$ transform as a 1-cycle. 
In case the Ricci tensor $R_{\ucA\cA}$ is proportional to the 
metric:  
\(
R_{\ucA\cA} = f^{2} g_{\ucA\cA}
\) 
(Einstein spaces), the \Kh\ potential can be written as
\(
\cK = f^{-2} \ln \det g
\labl{O.7.11}
\)
and the holomorphic functions are given by
\(
\cF_i = 1/(2f^{2})\, \cR_{i, \cA}^\cA
\).
Defining the Killing potentials $\cM_i(\bZ, Z)$ as 
\equ{
i \cM_i \equiv \cK_{,\cA} \cR_i^\cA - \cF_i = 
 -\cK_{,\ucA} \bcR_i^\ucA + \bcF_i,
\labl{O.7.2}
} 
with the second identity following from eq.\eqref{O.7}, one 
observes that the Killing potentials $\cM_i$ are real functions.
The Killing vectors $\cR^\cA_i$ can be obtained from them by
\equ{
i \cM_{i,\ucA} = g_{\ucA \cA} \cR^\cA_i, \qquad
i \cM_{i,\cA} = -g_{\ucA \cA} \bcR^\ucA_i.
\labl{O.7.3}
} 
They transform under the isometries in the adjoint representation
\equ{
\gd_{i} \cM_{j} = -\gd_{j} \cM_{i} = f_{ij}^{\;\;\;k} \cM_{k}.
\labl{O.7.4} 
} 
When the isometry group is semi-simple, all geometrical objects 
of the \Kh\ manifold can be expressed in terms of Killing potentials 
\ct{Aoyama:1986he, Aoyama:1986fj}.

If part of the internal symmetries are local, the partial 
derivatives in eq.\eqref{O.1} and in the \Kh\ covariant derivative 
$D_\gm$ have to be replaced by gauge covariant ones given by
\equ{
\barr{ll}
\der_\gm Z^\cA & \lra D_\gm Z^{\cA} = \der_\gm Z^\cA - 
A^i_\gm \cR^\cA_i, \\
& \\
D_\gm \gps_L^\cA & \lra \cD_\gm \gps_L^\cA = \der_\gm \gps_L^\cA  
- A^i_\gm \cR^\cA_{i, \cB} \gps_L^\cB + D_{\gm} Z^{\cC} 
 \gG_{\cC\cB}^{\cA} \gps^{\cB}, 
\earr
\labl{O.8}
}
where $A^i_\gm$ are the gauge fields corresponding to the 
local symmetries. They are components of the vector multiplets $V^i = 
(A^i_\gm, \gl^i, D^i)$, with $\gl^i$ representing the gauginos 
and $D^i$ the real auxiliary fields.

After the introduction of the gauge fields in the lagrangean 
\eqref{O.1} via the covariant derivatives \eqref{O.8}, the 
$\gs$-model itself is not invariant under supersymmetry 
transformations anymore. This is resolved by adding the terms 
\begin{equation}
\gD \cL_\cK \,=\, 2\, g_{\ucA \cA} 
\left( \bcR_i^{\ucA} \bgl_R^i \gps_L^\cA 
+ \cR_i^\cA \bgps_L^\ucA \gl_R^i \right) 
- D^i \lh \cM_i + \gx_i \rh 
\label{O.11} 
\end{equation} 
to the lagrangean \eqref{O.1}, including Fayet-Illiopoulos 
terms if applicable. 

The kinetic terms for these vector multiplets 
take the form \ct{Cremmer:1983en, Kugo:1983mr} 
\equ{
\cL_{f} \,=\, 
\left[  f_{ij} W^i(V)\, W^j(V) \right]_F = 
\frac 12 f_{ij}  
\left( - \bgl_R^i \lrset \sDer \gl_R^i - \frac 12 F^{i-}\cdot F^{j-} 
+ \frac 12 D^i D^j \right) 
\non \\ \non \\
+ \frac 12 f_{ij, \cA} \left( - \gs\cdot F^{i-} +i D^i \right)
 \bgps_R^\cA \gl_L^j 
- \frac 14 f_{ij,\cA} H^\cA \gl_R^i \gl_L^j 
\labl{O.9}
\\ \non \\
+ \frac 14 f_{ij, \cA\cB} (\bgps_R^\cA \gps_L^\cB)\, 
(\bgl_R^i \gl_L^j) + \mbox{h.c.},
\non
} 
where the $f_{ij}$ are chiral superfields transforming covariantly 
under the group of isometries; for example, they can be  
holomorphic functions of the chiral superfields $\gS$. The  
anti-selfdual field strength is defined as $F^{i-}_{\gm\gn} = \frac  
12 \lh  F^{i}_{\gm\gn} - \tilde F^{i}_{\gm\gn} \rh$. The covariant 
derivative acting on the gauginos is defined in the adjoint 
representation. The standard form of the function $f_{ij}$ is
$f_{ij}(Z) = \gs(Z) \get_{ij}$ where $\get_{ij}$ is the Killing 
metric defined from the structure coefficients 
\equ{ 
-2 \get_{ij} = -2 C_A \gd_{ij} = f_{ik}^{\;\;\;l} f_{jl}^{\;\;\;k},
\labl{O.9.1}
}
and $\gs(Z)$ is an invariant holomorphic scalar coefficient. The 
indices $i,j,\ldots$ run over the gauged part of the isometries 
and $\gd_{ij}$ is the Killing metric normalized by the Casimir 
$C_A$ of the adjoint. When a direct product group of subgroups 
is gauged, there are as many different coefficients $\gs^{(i)}$ 
as there are subgroups. The real parts of the coefficients: 
$\Re\, (C_A \gs^{(i)})$, can be interpreted as coupling constants 
$1/(g^{(i)})^2$.

From the covariance of the Killing potentials, eq.(\ref{O.7.4}), 
it is obvious that in non-supersymmetric models one can write
down a general class of non-minimal kinetic terms for the gauge 
fields, with $f_{ij}$ of the form
\equ{ 
f_{ij} = \gs \get_{ij} + \gr \cM_i \cM_j.
\labl{O.9.2}
} 
Here the coefficients $\gs$ and $\gr$ must be scalars under the 
internal symmetries. The inverse of $f_{ij}$ can be obtained if its
determinant is non-zero; this happens in particular if the Killing 
metric $\get_{ij}$ is invertable and 
\(
\gs, \gs + \gr (\cM_k)^2 \neq 0.
\) 
The non-minimal kinetic terms of this type are more complicated than 
the ones usually discussed in the context of $N=1$ supersymmetry, as 
for $\gr \neq 0$ the $f_{ij}$ are non-holomorphic functions of the 
chiral goldstone-boson superfields. Supersymmetrization of these terms 
is possible, but only at the expense of introducing higher-derivative 
terms for the chiral multiplets. 

To see this, we recall that in terms of components the $W^i$ 
form a particular kind of chiral spinor-tensor multiplet, 
obtained from the vector multiplet $V$ by
\equ{
W^i_L(V) \,=\, \lh \gl^i_L, 
\frac 12 \Bigl( - \gs\cdot F^{i-} + i D^i \Bigr),
\sDer \gl^i_R \rh.
\labl{O.10}
} 
Now the $\gs$-model action itself is derived from a composite abelian 
vector multiplet $\cK = (B_{\mu}, \gL, D)$, with components defined as
\begin{equation} 
\begin{array}{lll}
B_{\mu} & = & i K_{,\ga} \der_{\mu} z^{\ga}\, -\, i K_{\uga} \der_{\mu}\, 
  \bz^{\uga}\, + \,2 g V_{\mu}^i M_i\, - \,2i g_{\uga\ga}\, \bgps_L^{\uga} 
  \gg_{\mu} \gps_L^{\ga}, \\
 & & \\
\gL_R & = & 2 i g_{\uga\ga} \sDer \bz^{\uga}\, \gps_L^{\ga}\, - 2i 
  g_{\uga\ga} \hat{H}^{\ga} \gps_R^{\uga}\, -\, 2 g \gl^i_R M_i, \\
 & & \\
D & = & \cL\, \equiv\, \cL_{\cK}\, +\, \gD \cL_{\cK}.
\end{array} 
\label{new2.0}
\end{equation} 
Here $\cL = \cL_{\cK} + \gD \cL_{\cK}$ is the gauged extension of  
the $\gs$-model lagrangean, as discussed previously; by construction
it is invariant under supersymmetry modulo a total derivative. 

To the vector multiplet $\cK$ corresponds a similar spinor-tensor 
multiplet, with components 
\begin{equation} 
W_L(\cK) = \lh \gL_L, \frac 12 \Bigl( - \gs\cdot T^{i-} + i \cL \Bigr), 
 \sder \gL_R \rh,
\label{new2.1}
\end{equation} 
where 
 
\begin{equation} 
\begin{array}{lll}  
T_{\mu\nu} & = & \der_{\mu} B_{\nu} - \der_{\nu} B_{\mu} \\ 
 & & \\
 & = & g_{\ucA\cA} \lh D_{\mu} \bZ^{\ucA} D_{\nu} Z^{\cA} - 
 D_{\nu} \bZ^{\ucA} D_{\mu} Z^{\cA}\rh \\
 & & \\
 & & +\, \cD_{\mu} \lh g_{\ucA\cA} \bar{\psi}_L^{\ucA} \gg_{\nu} 
  \psi_L^{\cA} \rh - \cD_{\nu} \lh g_{\ucA\cA} \bar{\psi}_L^{\ucA} 
  \gg_{\mu} \psi_L^{\cA} \rh - i g F_{\mu\nu}^i \cM_i.
\end{array} 
\label{new2.2}
\end{equation}
Then the supersymmetric extension of the terms (\ref{O.9.2}) is 
constructed by taking the direct analogue of eq.(\ref{O.9}): 
\begin{equation} 
\gD \cL_{non-min}\, =\, \left[ \gs \eta_{ij} W^i(V) W^j(V) \right]_F\, 
 +\, \left[ \gr W(\cK)^2 \right]_F. 
\label{new2.3}
\end{equation}  
Clearly, this involves the square of the contracted field strength 
$[ F^i_{\mu\nu} \cM_i ]^2$, but in addition there are higher-derivative 
terms for the components of the chiral superfields $\gS^{\cA}$.

Until now we treated all chiral multiplets $\gS^\cA =\lh \gF^\ga, 
\, \gPs^A \rh$ on the same footing; we now classify the chiral 
multiplets $\gF^\ga$ and $\gPs^A$ by their transformation 
properties under the isometries. The chiral multiplets 
$\gF^\ga = \lh z^\ga, \gps_L^\ga, h^\ga \rh$ transforming 
non-linearly into themselves under a part of the isometries 
are called {\em $\gs$-model} multiplets. The chiral multiplets 
$\gPs^A = \lh x^A, \gch_L^A, f^A \rh$ transforming linearly 
into themselves under all isometries, but possibly with 
$\gs$-model-field dependent parameters, are called {\em matter} 
multiplets. The transformations eq.\eqref{O.5.2} of $\gs$-model 
and matter multiplets take the form 
\equ{
\gd_i \gF^\ga  = R_i^\ga (\gF), \qquad
\gd_i \gPs^A  =  R^A_{i\, B}(\gF) \gPs^B
\labl{O.16}
} 
according to the definitions above.
The Killing vectors \eqref{O.16} for the $\gs$-model and matter 
multiplets satisfy
\equ{
R^\gb_{[i} R^\ga_{j],\gb} = f_{ij}^{\;\;\;k} R^\ga_k, 
\qquad
R^B_{[i\, C} R^A_{j] \, B} + R^\gb_{[i} R^A_{j]\, C, \gb} = 
f_{ij}^{\;\;\;k} R^A_{k\, C}.
\labl{O.17} 
} 
The components of the $\gs$-model multiplets $\gF^\ga$ transform 
according to \eqref{O.5.1} but with $Z^\cA$ replaced by $z^\ga$, 
etc. The transformation rules for components of the matter multiplets 
$\gPs^A$ are more involved
\equ{
\gd_i \gPs^A = \left\{
\barr{ll}
\gd_i x^A & = R^A_{i\, B} x^B, \\
&\\
\gd_i \gch^A_L & = R^A_{i\, B} \gch^B_L + R^A_{i\,B, \gb} x^B 
  \gps^\gb_L, \\
& \\
\gd_i f^A & = R^A_{i\, B} f^B 
+ R^A_{i\,B, \gb} \lh x^B h^\gb - 2 \bgch^B_R \gps^\gb_L \rh 
- R^A_{i\, B, \gb\gg} x^B \bgps^\gb_R \gps^\gg_L.
\earr
\right.
\labl{O.21}
} 
Notice that the chiral matter fermions $\gch^A_L$ do not transform 
into themselves if the transformations $R^A_{i\, B}$ depend 
on the $\gs$-model fields. In section \ref{II} the chiral matter 
fermions are redefined such that they transform covariantly, 
see eq.\ \eqref{II.12} in that section.

Below we give a number of examples of matter multiplets and 
construct their \Kh\ potentials. Because the transformation rules 
for the complex matter scalars $x^A$ are linear in themselves, 
it follows that \Kh\  potentials for the matter multiplets are 
invariant unless the \Kh\ potential is a sum of holomorphic 
functions of these matter fields and their complex conjugates 
already\footnote{Of this trivial fact we make use later, see 
eq.\eqref{O.26.1}}. 

Given the Killing vectors $R_i^\ga$, the \Kh\ potential $K_\gs$ 
and hence the metric $g_{\gs\, \bga\ga}$ for the $\gs$-model 
multiplets, it is straightforward to give explicit examples 
of the transformations of matter multiplets.
By noticing that the metric $g_{\gs}$ defines an invariant 
line element on the \Kh\  manifold by
\(
d s^2 = d\bz^\uga g_{\gs\, \uga\ga}  d z^\ga,
\)
it follows that scalar fields $x^\ga$ which transform as 
differentials
\equ{
\gd_i x^\ga = R^\ga_{i, \gb}(z) x^\gb,
\labl{O.23}
} 
have an invariant \Kh\ potential given by 
\equ{
K_1 (\bx, x; \bz, z) = \bx^\uga g_{\gs\, \uga \ga} x^\ga.
\labl{O.24}
} 
With the subscript $1$ we indicate that this is the coupling of 
a rank one tensor (a vector) to the $\gs$-model. 
The complex scalar $x^\ga$ can be part of a chiral multiplet 
$\gPs^\ga = (x^\ga, \gch_L^\ga, f^\ga)$. 
Its transformation rules are given by equations \eqref{O.21} 
when $A, B$ are replaced by $\ga, \gb$. By taking tensor products 
of $n$ such vectors one can built a rank $n$ tensor chiral 
multiplet which is coupled to the $\gs$-model. It is possible to 
construct irreducible representations of the linear isometries 
by (anti)-symmetrizations and by taking traces. This construction 
is called covariant matter coupling \cite{vanHolten:1985xk, 
Clark:1985qy, GrootNibbelink:1998tz}.

It is also possible to couple a singlet chiral superfield 
$\gO = \lh s, \gch_L, \mbox{f} \rh$  non-trivially to a 
$\gs$-model \cite{GrootNibbelink:1998tz}. The singlet chiral 
multiplet $\gO$ transforms as
\equ{
\gd_i \gO = - f^2 F_i(\gF) \gO.
\labl{O.26}
} 
which forms a representation, see eq.\eqref{O.7.1}.
Note, that as the matter fields transform linearly into 
themselves, we have $\cF_i(\gF, \gPs) = F_i(\gF)$. 
We have taken the scalar component $\gO$ dimensionless, as
is convenient for the applications of $\gO$ later.
A covariant \Kh\ potential for this singlet $\gO$ is 
given by
\equ{
K_\gO = f^{-2} \ln \lh \bgO \gO \rh,
\labl{O.26.1}
}
which transforms opposite to the \Kh\ potential $K_\gs$ of the 
$\gs$-model fields. The components of the multiplet $\gO$ are 
non-propagating, as $K_\gO$ is a sum of a holomorphic and an 
anti-holomorphic function. One can use this {\em compensating} 
multiplet $\gO$ to rescale other matter multiplets so as to assign 
arbitrary $U(1)$ charges $q^{(A)}$ to them \ct{GrootNibbelink:1998tz}.
Indeed, let $\gPs^A$ be a set of matter multiplets described by a 
\Kh\ potential 
\(
\bgPs^\bA g_{\bA A} \gPs^A
\).
Define the rescaled multiplets ${\gPs'}^A$ by 
\(
{\gPs'}^A = \gO^{- q^{(A)}} \gPs^A
\).
Their transformation rules become
\equ{
\gd_i {\gPs'}^A  =  R^A_{i\, B}(\gF) {\gPs'}^B + q^{(A)}  
  f^2 F_i(\gF) {\gPs'}^A
\labl{O.30}
} 
and their \Kh\ potential has to be modified to 
\equ{
{\bgPs}^{\prime\bA} {g'}_{\bA A} {\gPs}^{\prime A} = 
{\bgPs}^{\prime\bA} g_{\bA A} {\gPs}^{\prime A} e^{- q^{(A)} f^2 K_{cov}}.
\labl{O.30.1}
} 
The \Kh\ potential $K_{cov}$ denotes any \Kh\ potential  
transforming in the same way as $K_\gs$. The case where 
$K_{cov} = K_\gs$ was discussed in ref.\ct{GrootNibbelink:1998tz}.
The numbers $q^{(A)}$ are arbitrary real numbers and may be fixed 
by dynamical considerations, like anomaly cancellation\footnote{
This works because one can define covariant matter fermions $\hat 
\gch^A_L$ using \eqref{II.12} transforming as 
\(
\gd_i \hat \gch^A_L = R^A_{i\;B} \hat \gch^A_L
\), 
as we show in section \ref{II}.
} 
\ct{GrootNibbelink:1998tz}. 

We finish this section by fixing the notation for the general 
considerations below. In the following we denote the \Kh\ potential 
for all physical $\gs$-model multiplets $\gF$ by $K_\gs(\bgF, \gF)$, 
and the \Kh\ potential for all physical matter multiplets $\gPs^A$ 
by $K_m(\bgPs, \gPs; \bgF, \gF)$. The matter fields $\gPs^A$ 
residing in $K_m$ may be rescaled by some power of $\gO$.   
$K = K_\gs + K_m$ is the sum of these two \Kh\ potentials. 

In the discussion of superpotentials \eqref{O.3}, it is often 
convenient to introduce a {\em compensating} superpotential 
$w(\gS)$: a dimensionless composite chiral superfield which 
transforms as 
\(
\gd_i w = q f^2 F_i w
\)
under the internal symmetries, with $q$ a real number\footnote{
Such a compensating superpotential $w$ can always be constructed: 
for instance add two physical singlets $S_+$ and $S_-$ with 
opposite charge to cancel any anomalies, and consider $w = f S_+ $.}.
With such a holomorphic function $w$, an {\em invariant} \Kh\ 
potential can be defined in terms of the physical fields only
\equ{
\cK(\bgS, \gS) =  K(\bgS, \gS) - \frac 1 {q f^2} \ln |\cW(\gS)|^2.
\labl{O.31} 
} 
Here the {\em covariant} superpotential $\cW$ is defined by 
\equ{
\cW(\gS) = f^3 w(\gS) W(\gS),
\labl{O.32}
} 
combining the {\em invariant} superpotential $W$, as in eq.\ 
\eqref{O.3}, with the {\em compensating} superpotential $w$ introduced 
above. Observe, that with a compensating singlet $\gO$ one can 
not make more general superpotentials than with physical 
multiplets alone, as $\gO$ can always be integrated out. 

\section{Non-linear chiral multiplet coupled to supergravity} \label{VII} 

We now turn to the coupling of gauged chiral multiplets to 
supergravity, as discussed for example in 
\ct{Cremmer:1983en,Kugo:1983mr}, generalized to include 
(holomorphic) non-linear gauge trans\-formations. This coupling 
to supergravity has also been discussed in \cite{Bagger:1983ab} 
using the superspace formalism \cite{Bagger:1990qh}. A related 
approach using \Kh\ superspace \cite{Binetruy:1990mp} can be 
found in \cite{Grimm:1990fp}. Besides presenting a review of 
this coupling, the main purpose of this section is to relate 
it to the rescaling of the matter multiplets we discussed in 
the previous section. We make the same distinction between 
$\gs$-model multiplets $\gF^\ga$ and matter multiplets $\gPs^A$ 
as in section \ref{O}. Of the latter the compensating singlet 
$\gO$ plays a crucial role in the locally supersymmetric case 
as well. The matter fields $\gPs^A$ are initially assumed to 
transform covariantly; rescalings by powers of $\gO$ come out 
naturally, as we show below. 

As was discussed in ref.\ \cite{Cremmer:1983en} an elegant way 
of coupling chiral multiplets to supergravity goes as follows: 
first couple the chiral multiplets to superconformal gravity, using 
a compensating multiplet $\gO$. By fixing a set of gauges involving 
this compensating chiral multiplet $\gO$ the superconformal algebra 
is reduced to the super-Poincar\'e algebra. On a chiral multiplet 
$\gS = (Z, \gps_L, H) \in \{\gF^\ga, \gPs^A, \gO \}$ the local 
superconformal algebra with transformations \(\gd = \gd_Q(\ge) + 
\gd_S(\get) + \gd_D(\gl) + \gd_A(\gth)\) is realized by
\equ{
\barr{lll} 
\gd Z & = & \bge_R \gps_L + \go (\gl - \frac{i}{3}\, \gth) Z  \\ 
& & \\ 
\gd \gps_L & = & \frac{1}{2}\, \lh D\Slashed  Z \ge_R + H \ge_L 
\rh + \go Z \get_L  + \left[ \lh\go + \frac{1}{2}\rh \gl + i  
\lh \frac{1}{2} - \frac{\go}{3} \rh \gth \right] \gps_L, \\  
& & \\ 
\gd H & = & \bar{\ge}_L \lh D\Slashed \gps_L - \gl_R^i  
 \cR_i(Z) \rh + 2 (1-\go) \bar{\get}_R \gps_L \\ 
 & & \\ 
 & & + \left[ \lh\go + 1\rh \gl + i \lh 1 - \frac{\go}{3}\rh   
\gth \right] H. 
\earr 
\labl{VII.5}
} 
Here $(\gl, \gth)$ are the parameters of local scale and chiral 
$U(1)$ transformations, whilst the spinors $(\ge, \get)$ parameterize 
local $Q$- and $S$-supersymmetry transformations, respectively. 
Furthermore $\go = (\go^{(\ga)}, \go^{(A)})$ denote the Weyl-weights 
of the chiral multiplets; the Weyl-weight of $\gO$ is taken to be 
$\go^{(\gO)} = 1$. The special conformal boosts do not have to be 
considered here as their only role is to fix the Weyl gauge field 
$b_\gm$ to zero when we restrict to Poincar\'{e} supergravity. 
The covariant derivatives are superconformal derivatives with the 
non-linear gauge-covariantizations (\ref{O.8}) included. 

Under the internal symmetries the $\gs$-model fields $\gF^\ga$ and the 
matter fields $\gPs^A$ transform according to eqs.\ \eqref{O.16}. 
Generically this requires the conformal Weyl weights of the 
$\gs$-model bosons to vanish; formally this is derived by 
requiring the internal symmetries and the space-time symmetries 
to commute: 
\equ{
0 = [\gd_D, \gd_i] z^\ga  = \go^{(\gb)} R^\ga_{i,\gb} z^\gb -  
\go^{(\ga)} R^\ga_i \hspace{1em} \Rightarrow \hspace{1em}  
\go^{(\ga)} = 0, \forall \ga. 
\labl{VII.6.1}
} 
By a similar argument any additional $U(1)$ symmetries entering 
the theory should leave the $\gs$-boson fields inert as well. 
Furthermore, the Weyl weights of the matter multiplets in a 
single irreducible representation must all be equal. We can make 
them vanish as well by multiplying with an appropriate power of 
the compensator: 
\equ{
{\gPs'}^A = \gO^{-\go^{(A)}} \gPs^A \hspace{1em} \Rightarrow 
\hspace{1em} \go^{\prime\, (A)} = 0. 
\labl{VII.7}
} 
Clearly, the dimension of the physical fields (as opposed to the 
Weyl weight) is kept fixed by taking $\gO$ dimensionless. 
For later use in constructing invariant actions we demand that 
the compensating superfield $\gO$ transforms like a non-trivial 
singlet \eqref{O.26} under the internal symmetries as 
\equ{
\gd_i \gO = - \gk^2 F_i(\gF) \gO, 
\labl{VII.5.1} 
} 
with $F_i(\gF)$ having vanishing Weyl weight: $\go^{(F_i)} = 0$, 
but ---like for the \Kh\ potential itself--- the mass-dimension 
dim$[F_i]$ = 2.
Therefore we have introduced the inverse Planck scale $\gk = 1/M_{P}$.  
By eq.\eqref{O.26} this implies that the multiplet ${\gPs'}^A$ 
transforms under the internal symmetries as 
\equ{
\gd_i {\gPs'}^A  =  R^A_{i\, B}(\gF) {\gPs'}^B + \go^{(A)} \gk^2 
F_i(\gF) {\gPs'}^A. 
\labl{VII.13}
} 
which is precisely the form of equation \eqref{O.30} with
$q^{(A)} f^2 = \go^{(A)} \gk^2$. These charges were introduced more 
or less ad hoc in ref.\cite{GrootNibbelink:1998tz}, so as to cancel 
anomalies. From now on we assume that we have performed this 
rescaling to all the matter fields, therefore we drop the prime 
on the matter fields. 

With these results in mind we proceed in the standard way 
\cite{Cremmer:1983en, Kugo:1983mr} to construct invariant 
functions of the superfields and use the density formula 
for real superfields of Weyl-weight 2 and chiral superfields of 
Weyl-weight 3 to obtain superconformally invariant lagrangeans. 
Let $K$ be the \Kh\ potential for the $\gs$-model fields $\gF^\ga$ 
and the matter fields $\gPs^A$ which is covariant 
\equ{
\gd_i K = F_i + \bF_i. 
\labl{VII.8.1} 
}
One defines a dimensionless invariant \Kh\ potential $\cG$ by 
\equ{ 
e^\cG = \bgO \gO\, e^{\gk^2\, K(\bgF, \bgPs; \gF, \gPs)}. 
\labl{VII.8} 
}
$e^{\cG}$ is a real superfield with Weyl-weight 2 and is 
inert under all internal symmetries. Hence by using the density 
formula for a real Weyl-weight 2 superfield it follows that the 
lagrangean 
\(
\left[ e^{\cG} \right]_D = \left[\bgO \gO\, e^{\gk^2 
K(\bgF,\bgPs,\gF,\gPs)} \right]_D 
\) 
is invariant under superconformal and internal symmetries. For 
similar reasons the only Weyl-weight 3 $F$-term lagrangean one 
can write down is 
\(
\left[(\gO)^3 \cW^{3/\go}(\gF, \gPs)\right]_F 
\) 
where the covariant superpotential $\cW$ is a dimensionless 
holomorphic function of $\gF^\ga$ and $\gPs^A$. This lagrangean 
is inert under the internal symmetries provided that the 
superpotential transforms as 
\equ{ 
\gd_i \cW(\gF, \gPs) = \go \gk^2\, F_i(\gF) \cW(\gF, \gPs), 
\labl{VII.15} 
}
again with the Weyl weight and rescaling charge $q$ of eq.\
\eqref{O.31} related by $q f^2 = \go \gk^2$. The particular 
power $3/\go$ of the superpotential $\cW$ is required 
precisely to satisfy this transformation rule. By redefining 
the compensating multiplet as 
\equ{
\gO' =  \gO \cW^{1/\go}(\gF, \gPs), 
\labl{VII.15.1} 
} 
it is inert under all internal symmetries and the superpotential 
can be absorbed into the extended \Kh\ potential $\cK$ 
\cite{Cremmer:1983en} 
\equ{
e^{\cG} = \bgO' \gO'\, e^{\gk^2 \cK}, 
\labl{VII.16}
} 
where $\cK$ is given by eq.\ \eqref{O.31} with the above 
substitution $q f^2 \rightarrow \go \gk^2$. Therefore a 
superpotential for the physical fields necessarily transforms 
as in eq.\eqref{VII.15}. From now on we use the redefined singlet 
$\gO'$ of eq.\eqref{VII.15.1}, unless explicitly stated otherwise; 
therefore we drop the primes on $\gO$. Although we now consider 
non-linear internal symmetries the \Kh\ potential $\cK$ takes 
the same form as given in ref.\cite{Cremmer:1983en}. But 
because of this non-linear nature, the gauging of part of 
the internal symmetries leads to some modifications of the 
invariant lagrangean. These modifications only come from the 
$D$-terms and gaugino-matter coupling which we discuss 
below\footnote{All gauge couplings now involve the Killing 
vectors $\cR^\cA_i$ as well.}. The crucial part of the lagrangean 
in eq. (3.14) of ref.\cite{Cremmer:1983en} generalizes to 
\equ{
e^{-1}\, \gD \cL\, =\, \frac{1}{4}\, f_{ij} \, D^i D^j\, +
\frac{i}{2\gk^2}\, \lh e^{\cG} \rh_{,{\cal A}} {\cal R}^{{\cal A}}_i 
\left( D^i + i \bar \gPs_R \cdot \gg \gl^i_R \right)  \non \\ 
\labl{VII.18} \\ 
 + \frac{2}{\gk^2}\, \lh e^{\cG} \rh_{,\ubar {\cal A} {\cal A}} 
{\cal R}^{{\cal A}}_i \bar \gl^i_L \gch^{\ubar {\cal A}}_R + 
\mbox{h.c.} \non
} 
where ${\cal R}^{{\cal A}}_i$ are the Killing vectors defined for 
all the fields in the model. Notice that $\cR_i^\gO = 0$ since 
the compensating multiplet $\gO$ does not transform under the 
internal symmetries. These Killing vectors can be obtained from 
Killing potentials ${\cal M}_i$ defined by 
\( i \gk^2 {\cal M}_i = \lh e^{\cG} \rh_{,{\cal A}} 
{\cal R}^{{\cal A}}_i, \) 
using \eqref{O.7.3} and the vanishing of $\cF_i$, as $\cG$ is 
inert under the internal symmetries. This holds for the \Kh\  
potential $\cK$ as well, and the Killing  potential can be 
expressed as 
\begin{equation}  
\cM_i = e^\cG M_i = e^\cG\, \lh M_{\gs i} - iK_{m, \ga} R^{\ga}_i
 - i K_{m, A} \left\{ R_{iB}^A + \go^{(A)} F_i \gd_B^A \right\}\, 
  \gPs^B \rh. 
\label{new3.0}
\end{equation}
For example, with the matter terms of the form 
\begin{equation} 
K_m\, =\, \sum_{\Psi} \bar{\Psi}^{\uA} g_{\uA A} \Psi^A e^{- \gk^2 
  \go^{(A)} K_{\gs}}, 
\label{new3.1}
\end{equation} 
the Killing potentials can be expressed as \ct{GrootNibbelink:1998tz} 
\begin{equation} 
\begin{array}{lll}
{\cal M}_i\, =\, e^{\cG} M_i & = & \dsp{ e^{\cG} \left( M_{\gs i}\, 
 \lh 1 + \eta \gk^2 \sum_{\Psi} \go^{(A)} \bar{\Psi}^{\uA} g_{\uA A} \Psi^A 
 e^{- \go^{(A)} \gk^2 K_{\gs}} \rh \right. }\\
 & & \\ 
 & & \dsp{ \left. -\, i \sum_{\Psi} \bar{\Psi}^{\uA} R_{i\, \uA, A} \Psi^{A} 
      e^{- \go^{(A)} \gk^2 K_{\gs}} \right). }
\end{array} 
\label{VII.20} 
\end{equation}
The part of the lagrangean \eqref{VII.18} can be written in 
terms of Killing potentials as 
\equ{ 
e^{-1}\, \gD \cL\, =\, \frac{1}{4}\,  f_{ij} \, D^i D^j\, -\, 
\frac 12 \cM_i \left( D^i + i \bar \gPs_R \cdot \gg \gl^i_R   
\right) + 2i \cM_{i,\ucA} \bar \gl^i_L \gch^{\ubar {\cal A}}_R + 
\mbox{h.c.}. 
\labl{VII.23} 
} 

The total lagrangean in superconformal gravity of gauged 
non-linear isometries with matter couplings is given by 
\equ{ 
\cL = \frac{1}{\gk^2}\, [ \bgO \gO e^{\gk^2 \cK}]_D + 
 \frac{1}{\gk^3}\, [\gO^3]_F + [f_{ij}W^i W^j]_F. 
\labl{VII.17.1} 
}
To reduce the lagrangean \eqref{VII.17.1} to Poincar\'{e} 
supergravity with matter coupled to it, one has to perform a 
number of gauge-fixings \cite{Cremmer:1983en}. This can be 
done in a clever way \cite{Kugo:1983mr} by choosing 
\equ{
\barr{cccc}
D: & \bs {s} e^{\gk^2 \cK}  = 3, & A: & 
\Im\ {s} = 0, \\  
\\
S: & {\gch}_L = - \gk^2 s \cK_{,{\cal A}} \gps^{{\cal A}}_L, & 
  K_m: & b_\gm = 0, 
\earr 
\labl{VII.17} 
} 
using the components of $\gO = (s, {\gch}_L)$. 

We now briefly review the relationship between this setup and 
the formulation of refs.\ \cite{Bagger:1983ab, Binetruy:1990mp}. 
If one does not perform the redefinition \eqref{VII.15.1} the 
superconformal lagrangean reads 
\equ{ 
\cL = \frac{1}{\gk^2}\, [\bgO \gO e^{\gk^2 K}]_D + 
 \frac{1}{\gk^3}\, [\gO^3 \cW^{3/\go}]_F + [f_{ij}W^i W^j]_F. 
\labl{VII.17.2} 
} 
It is inert under all internal symmetries provided that the 
compensator $\gO$ transforms according to eq.\ \eqref{VII.5.1}. 
If one reduces to Poincar\'{e} supergravity by applying the gauge 
fixings \eqref{VII.17} with $\cK$ replaced by $K$ the results of 
refs.\ \cite{Bagger:1983ab} are obtained. The gauge fixings eq.\ 
\eqref{VII.17} in this situation are not invariant under the 
internal symmetry transformations; this can be compensated by 
a chiral rotation \cite{Bagger:1983ab} 
\equ{
\gd_i \psi  = i \frac 12 \text{Im} F_i  \gg_5 \psi, 
\labl{VII.17.3} 
} 
on all spinors $\psi$. One could also consider arbitrary 
holomorphic functions $F$ in eqs.\ \eqref{VII.8.1} and 
\eqref{VII.5.1} instead of the functions $F_i$, which are 
dictated by the isometries \eqref{O.16}. This is the basis 
of \Kh\ superspace \cite{Binetruy:1990mp} where the \Kh\ $U(1)$ 
transformations are gauged. Notice that the redefinition of the 
compensator \eqref{VII.15.1} is a special case of this. It is 
clear that the lagrangean \eqref{VII.17.2} is invariant under 
these transformations. This reflects the fact that in supergravity 
the \Kh\ potential $K$ and the superpotential $\cW$ are not 
independent.
            
In this article our primary concerns are the isometries of the 
$\gs$-model and the holomorphic functions $F_i$ they induce. 
Therefore we choose to work  with the invariant \Kh\ potential 
$\cK$ and the lagrangean \eqref{VII.17.1}, amounting to a 
specific gauge in \Kh\ supergravity. 
            
\section{Vacuum configuration}\label{XI} 

In the previous section the construction of (locally) 
supersymmetric lagrangeans for $\gs$-models with non-linear 
symmetries was discussed. We now make a first step in the 
analysis of the phenomenology of such models. The scalar 
potential $V$ is given here before the auxiliary fields are 
eliminated. This has the advantage that breaking of supersymmetry 
is encoded in the vacuum expectation values of the auxiliary 
fields: supersymmetry is broken iff at least one auxiliary field 
has a non-vanishing VEV. We do not consider any fermion condensates 
here. Combining the results of eqs.\ \eqref{O.1}, \eqref{O.3} and 
\eqref{VII.23}, the scalar potential for the auxiliary and physical 
scalars in Poincar\'{e}  supergravity reads 
\equa{
V =\, & V_D + V_F = - \frac 12 \Re f_{ij} D^i D^j - g_{\ucA \cA} 
\bH^\ucA H^\cA 
\labl{XI.1} \\ 
& \non \\ 
& + D^i \cM_i - \gk\inv\lh \cK_{,\ucA} \bH^\ucA + \cK_{,\cA} H^\cA  
\rh e^{\frac 12 \gk^{2} \cK}- 3 \gk^{-4} e^{\gk^{2}\cK}, \non 
} 
using the results of ref.\ \cite{Kugo:1983mr} together with 
the generalization for non-linear symmetries of section \ref{VII}. 
In this scalar potential the \Kh\ potential is given by 
eq. \eqref{O.31}.
If supersymmetry is 
unbroken, the gravitino may still have a non-vanishing mass 
\equ{ 
\gk\inv e^{\frac 12 \gk^{2} \cK} \bgps_{\gm} \gs^{\gm\gn} 
\gps_{\gn} =\gk \inv |\cW|^{- \frac 1\go}e^{\frac 12 \gk^{2} K} 
\bgps_{R\gm} \gs^{\gm\gn} \gps_{L\gn} + \mbox{h.c.}, 
\labl{XI.2} 
} 
as this term, together with the negative cosmological last 
term of eq.\ \eqref{XI.1}, is supersymmetric. A vanishing 
gravitino mass is only possible if $-\frac 1\go > 0$ and the 
covariant superpotential vanishes in the vacuum. Because of 
the covariance of the superpotential $\cW$ the gravitino mass 
is linked to the breaking of internal symmetries, as 
\equ{ 
<\gd_i \cW>  = q f^2 <F_i> <\cW>. 
\labl{XI.4} 
} 
If $<\cW> \neq 0$, the symmetries for which $<F_i> \neq 0$ are 
broken. In particular the $U(1)$-factor of the linear isometries 
produces a constant function $F_{U(1)} = f^{-2} a$, hence this 
$U(1)$ is broken as soon as $<\cW> \neq 0$. Observe that the 
inverse is not necessarily true: when the gravitino mass 
vanishes ($<\cW> = 0$) symmetry breaking is not automatically 
ruled out.
            
In the absence of fermion condensates the equations of motion 
of the auxiliary fields in the vacuum are 
\equa{ 
\Re f_{ij} D^j  = &  \cM_i, \non \\ 
& \labl{XI.5} \\ 
\bH^{\ucA} g_{\ucA \cA} = & - \gk^{-1} \left(  \cW K_{,\cA} - 
\frac 1{q f^2} \cW_{,\cA}  \right) | \cW |^{- \frac 1\go} \cW^{-1} 
e^{\frac 12 \gk^2 K}. \non 
} 
If the metrics $\Re f_{ij}$ and $g_{\ucA \cA}$ are invertable, 
the scalar potential can be written in hybrid form as  
\equ{ 
V = \frac 12 \Re f_{ij} D^i D^j  + g_{\ucA \cA} \bH^\ucA H^\cA 
- 3 \gk^{-4} | \cW |^{- \frac 2\go}e^{\gk^{2}K}, 
\labl{XI.6} 
} 
with $D^i$ and $H^\cA$ the solutions \eqref{XI.5}. Using eqs.\ 
\eqref{XI.5} and \eqref{XI.6} the following phenomenological 
picture emerges. First of all observe that if 
\equ{ 
- \frac {\gk^2}{q f^2} =  -\frac 1\go < 1 \qquad 
(\exists \cW_{,\cA} \neq 0) 
\labl{XI.7} 
} 
the scalar potential diverges for $\cW = 0$ unless for all 
$\cA$ the first derivative of the superpotential vanishes as 
well. In that case the scalar potential still diverges if 
\equ{ 
-\frac 1\go < 0\qquad (\forall \cW_{,\cA} = 0). 
\labl{XI.8} 
} 
In this case the scalar potential can diverge to $\pm$ infinity 
depending on the details of the \Kh\ potential. (In the case where 
not all $\cW_{,\cA} = 0$ the potential always diverges to $+$ 
infinity, as $|\cW_{,\cA}|^2$ is always positive in that situation.) 
            
We now discuss the consequences of the analysis above. If 
$-\frac 1\go > 1$, the $\gs$-model cut-off is in general bigger 
than the Planck-scale. In this case $<\cW> = 0$ but the derivatives 
of $\cW$ do not have to vanish. The condition $<\cW> = 0$ may give 
rise to additional internal symmetry breaking. In the situation 
$1 \geq - \frac 1\go > 0$ the Planck scale may be much bigger 
than the $\gs$-scale, and not only $<\cW> = 0$ but also all 
$<\cW_{, \cA}> =0$. This means that there are more restrictions 
on the VEVs of the scalars and hence there may be more symmetry 
breaking and/or more parameters are fixed. In this case all the 
auxiliary fields \eqref{XI.5} of the chiral multiplets vanish, 
therefore $F$-term supersymmetry breaking is not possible. 
The spontaneous supersymmetry breaking can therefore only occur 
if the auxiliary $D$-fields \eqref{XI.5} acquire  non-vanishing 
VEVs. Soft supersymmetry breaking masses can still arise because 
of non-renormalizable contributions. We show how this works out 
in practice in section \eqref{IX}, where the vacuum configuration 
of a Grassmann $\gs$-model with a standard model-like spectrum is 
discussed.
            
\section{Grassmannian Manifolds} \label{VI} 

In this section we illustrate the general constructions discussed 
above with the example of Grassmann manifolds. Considering a 
particular model which decribes quark doublets, we show that anomaly 
cancellation is possible when we extend it to a non-linear version 
of the standard model. A Grassmann manifold is a homogeneous 
space which is obtained as the coset $U_\get(m,n)/U(m) \times U(n) 
\simeq SU_{\get}(m,n)/SU(m) \times SU(n) \times U(1)$. 
The parameter $\get$ distinguishes the compact and non-compact 
case: the compact group $(S)U(m+n)$ has $\get = 1$ and the 
non-compact group $(S)U(m,n)$ has $\get = -1$. Note that the 
second expression $G/H$ for the coset manifold is obtained 
from the first one by cancelling a $U(1)$ factor between the 
numerator and the denominator. This $U(1)$ may still act on the 
fields in our models, where it then represents a central charge 
(it commutes with all generators of $G$). Refs.\ \cite{Ong:1983te, 
Mattis:1983wn, vanHolten:1985di} provide the \Kh\  potential for 
these models, which can be written as 
\equ{ 
K_\gs(\bar Q, Q) \,= \, \frac 1{\get f^2} \lh {\frak a} \, 
\tr_m \ln (g\inv) + {\frak b}\, \tr_n \ln (\tg\inv) \rh 
\labl{VI.1} 
} 
with the inverse metrics $g\inv$ and $\tg\inv$ 
\equ{ 
\lh g\inv \rh^i_{\; j} \, = \, \left[1 + \get f^2 Q  
\bar Q \right]^{i}_{\;j}, \qquad\lh \tg\inv \rh^{a}_{\; b} \, 
= \, \left[1 + \get f^2 \bar Q Q\right]^{a}_{\; b}. 
\labl{VI.8} 
} 
Here $f$ is the parameter with the dimension of inverse mass 
setting the scale; it gives the fields $Q$ their canonical 
dimension. Two traces have been introduced: $tr_m$ acts on 
$m\times m$-matrices and $\tr_n$ on $n\times n$-matrices. The 
superfield matrix $Q = (Q^{ia})$ has vector indices in both 
$SU(m)$ and $SU(n)$ and $\bar Q = (\bar Q_{ai})$ is its conjugate. 
We take the indices $i = 1,\ldots, m$ and $a = 1,\ldots n$. 
In subsection \ref{VI.sub1} we interpret $Q^{ia}$ as a chiral 
multiplet containing a quark-doublet. The two real constants 
${\frak a}$ and ${\frak b}$ obey ${\frak a}+{\frak b}=1$ and 
hence drop out of \eqref{VI.1} after evaluating the traces. 
The constant ${\frak c}$ defined by $mn {\frak c} = m{\frak a} 
- n {\frak b}$, which may be used to characterize the central 
charge, is therefore not fixed uniquely. The non-linear 
realization of the $U_\get(m,n)$ algebra on multiplets $Q$ 
and $\bQ$ takes the form 
\equ{ 
\barr{cc} 
\gd Q = R(Q) \, = & \frac 1f \ge +\get f  Q  \bar \ge Q  + 
  i M Q - i Q N + i(m+n)\gth_Y Q, \\ 
 & \\ 
\gd \bar Q = \bR(\bQ) \,=\, & \frac 1f\bar \ge + \get f\bar Q \ge  
  \bar Q  + i N \bar Q - i \bar Q M - i(m+n)\gth_Y \bar Q, 
\earr 
\labl{VI.2} 
} 
where $M$ ($N$) represents the matrix of infinitesimal parameters 
of $SU(m)$ ($SU(n)$), $\bge$ an $n \times m$-matrix and $\gth_Y$ 
is a real number. We also introduce a real parameter $\gth_C$ 
for the central charge, but by construction the goldstone fields 
$Q$, $\bQ$ are inert under the central $U(1)$. The Lie algebra 
corresponding to  the tranformation rules \eqref{VI.2} can be 
stated as 
\equ{ 
\barr{c} 
\barr{ll}
[Y, X^{ia}] = (m+n)  X^{ia}, \quad & [Y, \bX_{ai}] = - (m+n) 
 \bX_{ai}, \quad \\ 
& \\  
\,[U^k_l, X^{ia}] =  \gd^i_l X^{ka} - \dsp{ \frac 1m} \gd^k_l X^{ia}, 
 \quad & [U^k_l, \bX_{ai}] = - \gd^k_i \bX_{al} + \dsp{\frac 1m}  
 \gd^k_l \bX_{ai} , \quad \\  
& \\ 
\,[ V^c_d, X^{ia}] = -\gd^a_d X^{ic} + \dsp{\frac 1n} \gd^c_d X^{ia},  
\quad & [ V^c_d, \bX_{ai}] = \gd_a^c \bX_{di} - \dsp{\frac 1n} 
 \gd^c_d  \bX_{ai}, \quad \\ 
 & \\ 
\,[ U^i_j, U^k_l ] = \gd^i_l U^k_j - \gd^k_j U^i_l,   &  
\,[ V^a_b, V^c_d ] = \gd^a_d V^c_b - \gd^c_b V^a_d, 
\earr \\ 
\\ 
\barr{l} 
[ \bX_{ai}, X^{jb} ] = \get \lh \gd^b_a U^j_i - \gd^j_i V^b_a \rh  
+ \get  \dsp{\frac 1{mn}} Y \gd^j_i  \gd^b_a, 
\earr 
\earr 
\labl{VI.3.5} 
\labl{VI.3.3} 
} 
where $U, V, X, \bX, Y$ are the generators of $SU_\get(m,n)$. 
By adding the generator $C$ for the central $U(1)$ we complete 
this to a full set of generators for $U_{\get}(m,n)$. The 
generators $U$ and $V$ are taken anti-hermitean and $X$ and $\bX$ 
are each others hermitean conjugates. The $U^i_j$ span the subalgebra 
$SU(m)$ of $U_\get(m,n)$ and similarly the generators $V^a_b$ span 
the subalgebra $SU(n)$. The two $U(1)$-factors in $U(m)$ and $U(n)$ 
combine to form the charges $Y$ and $C$. On $Q^{ia}$ the generators 
$U$ ($V$) act via left (right) multiplication. For this reason 
the commutators involving $V$ differ from the commutators 
involving $U$ by a minus sign. (By a redefinition of $V$ 
this minus sign could be absorbed.) The inverse metrics 
\eqref{VI.8} transform under these symmetries as 
\equ{ 
\gd g\inv = H g\inv + g\inv H^\dag,\qquad \gd \tg\inv 
 = \tg\inv \tH + \tH \tg^\dag. 
\labl{VI.3.2} 
} 
Here the holomorphic matrix-valued functions 
\equ{ 
H = \get f Q \bge + i M + i n \gth_Y + i \gth_C, \qquad\tH 
 = \get f \bge Q - i N + i m \gth_Y - i \gth_C. 
\labl{VI.3.7} \labl{VI.3.1} 
} 
and their conjugates transform in the adjoint representation of 
$U_\get(m,n)$. Using \eqref{VI.3.2} it is easy to show that 
$K_\gs$ in eq.\eqref{VI.1} transforms as a \Kh\ potential 
\equ{ 
\gd K_\gs(\bQ, Q) = F(Q) + \bF(Q). 
\labl{VI.9.1} 
} 
As the functions $H$ and $\tH$ transform in the adjoint 
representation, so does the holomorphic function 
\equ{ 
F(Q) =  \frac 1{\get f^2} \left( {\frak a} \tr_m H +  
 {\frak b} \tr_n \tH \right)= \frac 1{\get f^2} \lh \get f 
 \tr_m  (Q \bge) + i mn \gth_Y + i mn {\frak c} \gth_C \rh. 
\labl{VI.3.8} \labl{VI.3} 
} 

Next we discuss matter coupling to the Grassmannian model. 
Let ${\frak R}(\bgS, \gS)$ and $\tilde {\frak R}(\bgS, \gS)$ 
be $m\times m$-, resp.\ $n\times n$-matrix-valued composite 
real superfields. They are called {\em left}, resp.\ {\em right}, 
covariant if they transform as 
\equ{ 
\gd {\frak R} = H {\frak R} + {\frak R} H^\dag, \qquad\gd 
\tilde{\frak R} = \tilde{\frak R} \tH + \tH^\dag \tilde{\frak R} 
\labl{VI.10.3} 
} 
under the $U_\get(m,n)$ isometries of the Grassmannian manifold. 
Invariant \Kh\ potentials for these real  composite superfields 
${\frak R}$ and $\tilde{\frak R}$ are provided  by 
\equ{ 
\tr_m (g {\frak R}), \qquad \tr_n  (\tg \tilde{\frak R}). 
\labl{VI.10.4} 
} 
By eqs.\eqref{VI.3.2} it follows that $g\inv$, resp. $\tg\inv$, 
are left, resp. right, covariant but the construction mentioned 
above gives trivial results for these examples. To obtain 
non-trivial results, consider the chiral multiplets $L^i$ and 
$D^a$ which transform under $U_\get(m,n)$ by left, resp.\ right, 
multiplication 
\equa{ 
\gd L = & H L  = (\get f Q \bar \ge + i M+ i n \gth_Y + i \gth_C) 
 L, \non \\ 
 & \labl{VI.6} \\ 
\gd D = \ & D \tH = D ( \get f \bar \ge Q - i N + i m \gth_Y -  
 i \gth_C ). \non 
} 
We will later interpret $L$ and $D$ as chiral superfields 
containing the left-handed lepton doublets and charge conjugate 
of the right-handed $d$-quark. It follows that $L$ has a $Y$ 
charge $n$ and central $C$ charge $1$ and $D$ has $Y$ charge $m$ 
and central $C$ charge $-1$. However this interpretation does not 
work directly as for $m = 3$, $n = 2$ the $Y$ charges of $L$ and 
$D$ with respect to $Q$ do no t reproduce the standard hypercharges 
$Y_w$. Notice that $(L \bL)_i^j$ and $(\bD D)_a^b$ are left-, resp.\ 
right-, covariant composite superfields and hence from the expressions 
\eqref{VI.10.4} the \Kh\ invariants can be constructed 
\equ{ 
\bL g L \qquad \mbox{and} \qquad D \tg \bD. 
\labl{VI.9} 
} 
By taking tensor products of multiplets which transform like 
$L$ and $D$, one can obtain higher rank $U(m)\times U(n)$ tensors 
chiral multiplets. 
            
As the function $F$ defines a cycle, transforming with the 
structure constants of the gauge group, see eq.(\ref{O.7.1}),
we can use \eqref{O.26} to couple a multiplet $\gO$ which is 
a singlet under the semi-simple part of the unbroken symmetries 
to a Grassmannian manifold by 
\equ{ 
\gd \gO = \get f^2 F(Q) \gO  = \lh\get f 
\tr_m  (Q \bge) + i mn \gth_Y + i mn {\frak c} \gth_C\rh \gO. 
\labl{VI.4} 
} 
For later convenience, we take $\gO$ dimensionless. The 
rescalings with this singlet changes the $Y$ charge as well 
as the central charge $C$. Notice that we can introduce another 
singlet $\gO'$ with the same transformation rules as $\gO$ but 
with a different value for the central charge ${\frak c}'$, 
as the choice of parameters ${\frak a}$ and ${\frak b}$ in 
eq.\eqref{VI.1} is not unique. Therefore we can define two 
independent non-trivial singlets $\gO_Y$ and $\gO_C$ which 
transform as 
\equ{ 
\gd \gO_Y =  \lh \get f \tr_m  (Q \bge) + i mn \gth_Y \rh \gO_Y, 
\qquad\gd \gO_C = i mn \gth_C \gO_C,
\labl{VI.3.8a} 
} 
where we set the central charge ${\frak c}$ of $\gO_C$ to unity. When 
rescalings with $\gO_Y$ are performed, one needs to modify  
the metrics \eqref{VI.8} because this rescaling also generates 
additional non-linear transformations. For rescalings with 
$\gO_C$ this is not the case; it can in principle be applied 
to all multiplets. In the following we discuss the effects of 
rescalings on matter with a general singlet $\gO$ only, as 
rescalings with $\gO_Y$ or $\gO_C$ are just particular examples 
of this. 
            
Any given chiral multiplet, for example $L$, can be rescaled 
by a (non-physical) singlet $\gO$ to  \( L' = \gO^l L,\) 
which transform as 
\equ{ 
\gd L' = \lh l \, \get f^2 F +  H \rh L' 
\labl{VI.9.2} 
} 
using the transformation \eqref{VI.4} of the singlet $\gO$. 
In this way the right charges can be assigned to multiplets 
allowing for specific physical applications. The additional 
terms in the transformation rule for $L'$ have to be compensated 
in the \Kh\ potential. Again let ${\frak R}(\bgS, \gS)$ and 
$\tilde{\frak R}(\bgS, \gS)$ be left and right covariant real 
composite multiplets. Using eqs. \eqref{VI.3}, \eqref{VI.10.3} 
and \eqref{VI.9.2} a left covariant composite real 
superfield is constructed for $L'$ by 
\equ{ 
{\det}^{-l {\frak a}} {\frak R}\, {\det}^{-l {\frak b}} 
\tilde{\frak R}\ \;  L' \bL' \,= \,  e^{-l \lh {\frak a} 
\tr_m \ln {\frak R} + {\frak b} \tr_n \ln \tilde{\frak R} \rh} 
\,L' \bL' 
\labl{VI.10.6} 
} 
and hence using \eqref{VI.10.4} a \Kh\ invariant for $L'$ is 
obtained. Notice that this is an example of eq. \eqref{O.30.1} with 
\( 
K_{cov} = {\frak a} \tr_m \ln {\frak R} + {\frak b} 
\tr_n \ln \tilde{\frak R} 
\). 
If one takes $g\inv$ and $\tg\inv$ for the composite superfields 
${\frak R}$ and $\tilde {\frak R}$ then one obtains from this 
construction the invariant 
\equ{ 
\bL' g L'\, e^{- l \get f^2 K_\gs} 
\labl{VI.10.5} 
} 
by eq.\eqref{VI.1}. Of course a similar construction works for 
$D$ as well. After rescaling $L$ by $l$ and $D$ by $d$ such that: 
\equ{ 
\gd L = (H + l \get f^2 F) L, \qquad\gd D = D (\tH + d \get f^2 F), 
\labl{VI.3.11} 
} 
the generalizations of the \Kh\, invariants \eqref{VI.9} are 
given by 
\equ{ 
K_L = \bL g^{(L)} L, \qquad K_D = D \tg^{(D)} \bD, 
\labl{VI.10.7} 
} 
with the modified metrics 
\equ{ 
g^{(L)} = e^{-l\get f^2 K_\gs}\, g, \qquad\tg^{(D)} 
 = e^{-d\get f^2 K_\gs}\, \tg. 
\labl{VI.10.8} 
} 
We next turn to discuss the Killing potentials. We denote all 
Killing potentials $\cM_i$ collectively as \(\cM = \gth^i \cM_i,\) 
where $\gth^i$ stands for the parameters of the isometries. We 
first focus on the part $\cM_\gs$ of the Killing potential 
depending on the $\gs$-model fields $Q$ and $\bQ$ only; afterwards 
the matter contribution $\cM_m$ is examined. The complete Killing 
potential is given by $\cM = \cM_\gs + \cM_m$. Both the $\gs$-model 
and matter Killing potentials can be written conveniently in terms 
of the matrices  
\equa{  
\gD = &  R^{(ia)} \lh g\inv\rh_{,(ia)} g - H \non \\  
= & - i \gth_C + i \gth_Y \lh (m+n) \get f^2 Q \bQ g  - n \rh  
- i M g \non \\  
& \qquad - i \get f^2 Q N \bQ g + \get f \lh \ge \bQ - Q \bge \rh g,  
\non \\  
& \labl{VI.16} \labl{VI.15} \\  
\tilde \gD = &  \tg \lh \tg\inv\rh_{,(ia)} R^{(ia)} - \tH \non \\  
= & + i \gth_C + i \gth_Y \lh (m+n) \get f^2 \tg \bQ Q - m \rh+ i  
\get f^2 \tg \bQ M Q \non \\  
& \qquad+ i \tg N + \get f\tg \lh \bQ \ge - \bge Q \rh, \non  
}  
under some mild assumptions as we see below.  
Here we have used the index notation $(ia)$ to emphasize that  
this index refers to the superfield $Q^{ia}$. 
Using eq.\eqref{O.7.2} for the $\gs$-model fields $Q$ and $\bQ$ 
we find that their Killing potential can be written as  
\equ{  
i \cM_\gs = K_{\gs, (ia)} R^{(ia)} - \frac 1{q\get f^2}  
\cW\inv \gd \cW= \frac {{\frak a}}{\get f^2} \tr_m \gD +  
\frac {{\frak b}}{\get f^2} \tr_n \tilde \gD.  
\labl{VI.17} \labl{VI.17.1}  
}  
Notice that $\cW\inv \gd \cW$ plays the role of $F$ and that  
the covariant superpotential $\cW$ in combination with the  
$\gs$-model \Kh\ potential forms an invariant.  
 
To discuss the Killing potential due to matter fields in some  
generality we introduce some further notation. We discuss only  
the rescaled matter field $L$ here, as it is easy to generalize  
our discussion to the matter field $D$ and tensor products.  
Define the $m\times m$ matrix real composite superfield  
\([L \bL]_i^j =  (g^{(L)}\, L)^j \bL_i\) where $g^{(L)}$ is  
the rescaled metric defined in eq.\eqref{VI.10.8}. Notice that  
$[L \bL] g\inv$ is a left covariant real composite superfield,  
hence by \eqref{VI.10.4} we obtain the \Kh\, invariant:  
$\tr_m [L \bL] = K_L$. From now on we assume that the matter  
\Kh\, potential $K_m$ can be written entirely in terms of  
matrices like $[L \bL]$. As $K_m$ is an invariant \Kh\,
function, one can define the Killing potential for the matter  
field $L$ as  
\equ{  
i \cM_L = \tr_m \left[ K_{m, [L \bL]}\Bigl(\gd Q^{ia}  
( g^{(L)} )_{,(ia)}L \bL + g^{(L)} \gd L \bL\Bigr)\right].  
\labl{VI.18}  
}  
where $K_{m,[L \bL]}$ denotes the derivative of $K_m$ with  
respect to the matrix $[L \bL]$. This can be expressed in terms  
of $\gD$ and $\cM_\gs$ as  
\equ{  
i \cM_L = - \bL K_{m, [L \bL]} g^{(L)} \left(l \get f^2\,  
i \cM_\gs + \gD\right) L.  
\labl{VI.19}  
}  
The Killing potential $\cM_m$ due to all the different matter  
fields is a sum of Killing potentials like $\cM_L$. As the Killing  
potentials $\cM_\gs$, $\cM_m$ for the $\gs$-model fields and  
the matter fields are linear in $\gD$ and $\tilde \gD$, cf.\  
eq.\eqref{VI.17}, we can always express the full Killing potential  
as:  
\equ{  
i \cM = \tr_m \gD P+ \tr_n \tilde \gD \tP  
\labl{VI.20}  
}  
where the field dependent matrices $P$ and $\tP$ encode the  
details of the full \Kh\ potential. Using these matrices one  
can state the Killing potentials for the different symmetries  
of $U_\get(m,n)$ as  
\equa{  
\cM_C & = - \tr_m P + \tr_n \tP,\non \\  
\cM_Y & = \tr_m P \lh (m+n) \get f^2 Q \bQ g - n \rh \non \\  
& \qquad + \tr_n \tP \lh (m+n) \get f^2 \tg \bQ Q - m \rh, \non \\  
\cM_U & = - g P + \get f^2 Q \tP \tg \bQ,  
\labl{VI.21} \\  
\cM_V & = \tP \tg - \get f^2 \bQ g P Q, \non \\  
i \cM_X & = \get f \lh \bQ g P + \tP \tg \bQ \rh, \non \\  
i \cM_\bX & = - \get f \lh g P Q + Q \tP \tg  \rh. \non  
} 
In combination with the vector field strengths, also transforming  
according to the adjoint representation, we can now construct the  
invariant $\cM_{\mu\nu} = f^2 \cM_i  F_{\mu\nu}^{i}$. For example, 
if we gauge the full $SU_{\get}(m,n)$ (but without the central charge), 
the Killing potentials $\cM_{\gs i}$ of the pure $\gs$-model give 
\begin{equation} 
\begin{array}{ll}
\cM_{\mu\nu}(V,W,Z,\bZ,A) & =\, \dsp{  
 - if\, \mbox{tr}_m \left\{ g \lh F_{\mu\nu}(Z) \bQ + i f Q \bQ\,  
 (F_{\mu\nu}(V) + n F_{\mu\nu}(A)) \rh \right\} }\\  
 & \\ 
 & \dsp{ +\, if\, \mbox{tr}_n \left\{ \tg \lh F_{\mu\nu}(\bZ)  
 Q + if \bQ Q\, (F_{\mu\nu}(W) - m F_{\mu\nu}(A)) \rh \right\}. }\\ 
 & \\ 
 & -\, mn \eta\, F_{\mu\nu}(A).   
\end{array} 
\label{new.3} 
\end{equation}   
Here $(V_{\mu},W_{\mu},Z_{\mu},\bZ_{\mu},A_{\mu})$ are the vector  
fields for $SU(m)$, $SU(n)$, the off-diagonal generators of  
$SU_{\eta}(m,n)$, and $U(1)$, respectively. The kinetic terms for the  
gauge fields then can be constructed as  
\begin{equation}  
e^{-1}\, \cL_{gk}\, =\, -\gs\, \eta_{ij}\,   
 F_{\mu\nu}^{i} F_{\mu\nu}^{j}\, +\, \gr\, \left[  
 \cM_{\mu\nu} \right]^2\, + ..., 
\label{new.4} 
\end{equation}  
where the dots denote the supersymmetric completion. We observe,  
that in the case that all isometries are gauged (for $\eta = +1$),   
the unitary gauge is $Q = \bar{Q} = 0$, and therefore the higher-order  
scalar derivative terms are absent. As $\cM_{\gs}$ acquires a vacuum  
expectation value, it becomes constant and the non-minimal gauge-kinetic 
terms become of minimal type, but with a renormalized value of the 
$U(1)$ gauge coupling w.r.t.\ the gauge coupling of the other $SU(n+m)$ 
fields; after some rescaling 
we find   
\begin{equation}  
\begin{array}{lll} 
e^{-1} \cL_{gk}  & = & \dsp{ - \frac{\gs}{4}\, \lh \mbox{tr}_m
 F_{\mu\nu}(V)^2\, +\, \mbox{tr}_n F_{\mu\nu}(W)^2\, +\, 
 2 \mbox{tr}_m F_{\mu\nu}(Z) F_{\mu\nu}(\bZ) \rh\, }\\ 
 & & \\ 
 & & \dsp{ -\, (m+n)mn\, \frac{\gs}{4}\, \lh 1 + \frac{\gr}{\gs} 
 \frac{mn}{m+n})\rh F_{\mu\nu}(A)^2.} 
\end{array} 
\label{new.6} 
\end{equation}  
Note that the $D$-potential which accompanies the gauging induces 
mass terms for the vector bosons $(Z,\bZ)$. In the present normalization 
of the lagrangean the mass-term for the heavy gauge bosons is just 
$m_Z = 1/f$, but the physical masses then become to lowest order 
$M_Z^{phys} = 1/f \sqrt{\gs}$. 

\section{Grassmannian standard model}
\label{VI.sub1} 

We now turn to an example illustrating how one can cancel anomalies 
by adding rescaled matter multiplets. If we consider the case with 
$m=2$ and $n=3$ then the Grassmannian manifold may be the basis of 
an $SU(5)$ unification model with the standard model group 
$SU(2)\times U(1) \times SU(3)$ as the unbroken subgroup. We do 
not require the $U_\get(2,3)$ to be compact nor do we disregard the 
central charge $C$. In the standard model the field content is such 
that all possible anomalies cancel in each generation, 
consisting of a quark doublet $q^{\;}_L$, a lepton 
doublet $l_L$, quark singlets $d_L^c$ and $u_L^c$ and an electron 
singlet $e^c_L$. (The notions singlet and doublet here refer to 
$SU(2)$ representations.) If one wants to consider models with 
more generations, the simplest thing is just to take a number 
of copies of this structure.
Only in the quark doublet sector there will be a difference: an 
additional quark doublet $Q'$ is to be coupled covariantly to 
the $\gs$-model spanned by $Q$. We do not pursue multiple 
generation models here further. Finally we introduce a Higgs sector 
consisting of two $SU(2)$ doublets $H^\pm$ with opposite charge.
The introduction of the Higgses is necessary for the breaking 
of $SU(2)\times U_Y(1)$ to $U_{em}(1)$.

The hypercharges in the standard model are assigned so as to
produce anomaly cancelation. In the supersymmetric models the  
chiral fermion representations have to be completed to the chiral 
supermultiplets $Q^{ia}, L^i,  D^a,  U^a$, $H^\pm$ and $ E$. However 
if we use the standard coupling of matter multiplets to the 
Grassmann $\gs$-model we do not obtain the correct charge assignment.
\begin{table}
\begin{center}
\begin{tabular}{| l | l | c  c | c | r | l |}
\hline
Multiplet & Fermion & $Y$ & & $C$ & $Y_w$ & $k$ \\ \hline 
 $Q^{ia}$ & $q_L^{ia}$ & $n+m$ & 5 & $ {\frak c}_q$ & +1/3 & 0 \\ 
 $Q^{\prime ia}$ & $q_L^{\prime ia}$ & $n+m$ & 5 & ${\frak c}_q$ & 
 +1/3 & 0 \\ 
 $L^i$ & $l_L^i$ & $n$ & 3 & ${\frak c}_l$ & -1 & $l=-3$ \\ 
$H^{-i}$ & $h_L^{-i}$ & $n$ & 3 & ${\frak c}_-$ & -1 & $h^-=-3$ \\ 
$H^{+i}$ & $h_L^{+i}$ & $n$ & 3 & ${\frak c}_+$ & +1 & $h^+=+2$ \\ 
$D^a$ & $d_L^{c\,a}$ & $m$ & 2 & ${\frak c}_d$ & +2/3 & $d=4/3$ \\ 
$U^a$ & $u_L^{c\, a}$ & $m$ & 2 & ${\frak c}_u$ & -4/3 & $u=-11/3$ \\ 
$E$ & $e^c_L$ & 0 & 0 & ${\frak c}_e$ & 2 & $e= 5$ \\ 
$\gO$ & - & $mn$ & 6 &${\frak c}$& - & - \\ \hline 
\end{tabular} 
\caption{Grassmannian (matter) multiplets and their chiral fermion 
content. $Y$ is the canonical charge of the $\gs$ model and $Y_w$ 
denotes the hypercharge needed for anomaly cancelation within the 
standard model. These charges can be indentified if $Y = 15 Y_w$.
The number $k$ gives the rescalings with a singlet $\gO_Y$. 
$C$ is the central charge, which can be chosen differently for 
each (matter) multiplet, using the singlet $\gO_C$.} 
\label{tab.VI.1} 
\end{center} 
\end{table} 

In table \ref{tab.VI.1} the hypercharge $Y_w$ for the chiral 
multiplets containing the quarks and leptons is compared 
to the canonical charge $Y$ defined on the Grassmannian manifold, 
as e.g.\ obtained from the couplings in eqs.\ \eqref{VI.2} 
and \eqref{VI.6}. In the third column we have evaluated these 
$U(1)$ charges in the case of $U_\get(2,3)$ ($SU(5)$). In the fourth 
column we have given the central charges $C$ of the multiplets. 
As is obvious from this table the hypercharges $Y_w$ required in 
the standard model do not match the charges $Y$.
(For this to happen, we should have $Y = 15 Y_w$ for all fields.)
However from eq.\ \eqref{VI.4} we see that the singlet chiral 
multiplet $\gO_Y$ has $U(1)$ charge $mn = 6$ in the $U_\get 
(2,3)$ model. By employing the rescaling:
\(
\gPs^{(k)} = \gO_Y^k \gPs
\)
any chiral multiplet $\gPs$ can be given an additional charge 
$k\, mn$. In the last column we have given the powers 
($l, d, u, e, h^\pm$) to which the singlet has to be raised 
in order the find the right hypercharge assignment for 
the standard model. In a similar way the central charges $C$ 
may be ajusted to coincide with the $B-L$ quantum numbers. 

In the following we assume that we have performed the rescaling 
to the chiral multiplets as given in this table and hence we can 
state the \Kh\  potential. 
\equ{
K \,=\, K_\gs  + K_E + K_L + K_D + K_U + K_{H^+} + K_{H^-},
\labl{VI.11}
}

\nit
where $K_L$ and $K_D$ are defined in eqs.\eqref{VI.10.7} and 
$K_E$, $K_{H^\pm}$ and $K_U$ are defined in a similar fashion.

As fundamental compensating superpotentials we may take
\equ{
\begin{array}{ll}
w_{E\;} =   fE &\qquad (q_E = e = 5), 
\\ &  \\
w_{L-} =  f^2 \gve_{ij} L^i H^{-j} &  
 \qquad (q_{L-} = 1 + l + h^- = -5).
\end{array}
\labl{VI.22}
} 
Using the $SU(2)$ invariant $\gve$-tensor, it follows that 
$w_{L-}$ is a $SU(2)$ singlet. These compensating superpotentials 
transform as $\gd w = q \get f^2 F w$ where the numbers $q$ is 
given in the brackets in \eqref{VI.22}. The central charge 
${\frak c}_w$ of the compensating superpotential determines the 
central charge of the holomorphic functions into which the \Kh\ 
potential transforms:
\(
q mn {\frak c} = {\frak c}_w
\).
Combining these compensating superpotentials to another 
superpotential puts restrictions on the choice of $C$-charges 
of the superpotentials \eqref{VI.22}. For example 
\equ{
w =  a w_E + b (w_{L-})\inv,
\labl{VI.24}
} 
where $a$ and $b$ are complex constants, demands that 
\(
{\frak c}_{w_E} =  -{\frak c}_{w_{L-}}
\).
This in turn puts restrictions on the $C$-charges of the matter 
fields.

For the invariant superpotential $W$ we take a part of the standard 
model superpotential: 
\equ{
W = \ga + \gb E \gve_{ij} H^{-i} L^j - \gm \gve_{ij} H^{+i} H^{-j}.
\labl{VI.23}
} 
The first term $\ga$ is a constant with dimension of (mass)$^3$; 
the second term is the usual Yukawa coupling in supersymmetric 
models and the third term is the Higgs interaction. Notice that 
in this model there are no Yukawa interations for the quarks, as 
the quark doublet superfield $Q$ does not transform covariantly. 
Notice that the superpotential $W$ has a homogeneously vanishing 
central charge if 
\(
{\frak c}_e +  {\frak c}_l +  {\frak c}_- = 0 
\) 
and
\(
{\frak c}_+ +  {\frak c}_- = 0
\).
Also notice that these central charges can be choosen in accordance 
with the lepton number $L$. In that case we can take
\(
{\frak c}_+ =  {\frak c}_- = 0
\)
and 
\(
- {\frak c}_e = {\frak c}_l = 1
\).
The central charges of the quark multiplets ($D$, $U$) can be 
chosen to match the baryon number. However, this method 
does not apply to the left-handed quarks described by the 
$\gs$-model superfields $Q$

\section{Separation of submanifolds}
\label{II} 

If one considers the combined system of a non-linear $\gs$-model 
with additional matter coupling to it, the metric of that total 
system is in general not diagonal: one can have mixing between 
different representations of the symmetry algebra in the 
quadratic kinetic terms of the scalars and chiral fermions. 
This is carried over to the definition of propagators. 
If one knows that the theory is constructed out of several sectors, 
one would like to be able to assign to each sector a separate metric, 
without mixing between different sectors. This requires the metric 
to be block diagonal, with each block representing the metric 
of a different representation of the isometry group. With the 
machinery developed in the appendix this can be done elegantly 
without too much computational difficulty. 

We consider a \Kh\ manifold parametrized by the coordinates 
$Z^\cA = (z^\ga, x^A)$ and their conjugates. 
The method we follow generalizes the result of ref.\ 
\ct{Achiman:1985fh} where only quadratically coupled rank $1$ 
matter was considered:$ K_m = \bx^\uga g_{\uga \ga} x^\ga$, 
with the metric $g_{\uga ga}$ depending on $z^\ga$ and $\bz^\ga$. 
However, the starting point of this section is more general, 
allowing any \Kh\ potential $K$ of the $\gs$-model and matter 
fields. First we identify the $\gs$-model \Kh\ potential 
$K_\gs(\bz, z)$ and the matter \Kh\ potential $K_m(\bx, x; \bz,z)$ 
by
\equ{
K_\gs =\left. K\right|_{x= \bx = 0}, \qquad
K_m \equiv g_{\bx x} = K - K_\gs.
\labl{II.1}
} 
The notation $g_{\bx x}$ for the matter \Kh\  potential is very 
suggestive as it reduces to $g_{\bx x} = \bx^\uA g_{\uA A} x^A$ 
when matter is quadratically coupled. To take this analogy to the
case of quadratic matter coupling a bit further, we define
\equ{
g_{\bx A} = \cK_{,A}, \qquad
g_{\ubar  A x} = \cK_{,\ubar A}, \qquad
\labl{O.19}
} 
whilst the metrics for the matter and $\gs$-model fields are 
\equ{
g_{\uA A} \equiv \cK_{, \uA A}, \qquad
g_{\gs\, \uga\ga} \equiv K_{\gs, \uga\ga}.
\labl{O.20}
} 

To be able to use the method explained in the appendix, we 
first need to define the non-holomorphic transformation matrices 
$X^{\cA'}_\cA$ and $\bX^{\ucA'}_\ucA$. 
We do this by demanding that the transformations \eqref{I.5} 
block-diagonalize the metric of the combined system 
of $\gs$-model fields and matter and leave the metric for the 
matter fields unchanged. 
The metric of the combined system is
\equ{
g_{\ucA \cA} = \pmtrx{
g_{\gs\,\uga \ga} + g_{\bx x, \uga \ga} & g_{\bx A,\uga } 
\\
g_{\uA  x ,\ga} & g_{\uA  A}
},
\labl{II.7}
} 
where $g_{\gs\, \uga\ga}$ is the metric of the $\gs$-model without 
matter coupling. Then the appropriate transformation is given by 
the matrices
\equ{
X^{\cA}_{\cA'} = \pmtrx{
\gd^{\ga}_{\ga'} & 0 
\\[5pt]
- \gG^{A}_{x \ga'} 
&
\gd^{A}_{A'}
}
,\quad
\bX ^{\ucA}_{\ucA'} = \pmtrx{
\gd^{\uga}_{\uga'} &  
- \bgG^{\uA}_{\bx \uga'} 
\\[5pt]
 0  &  \gd^{\uA}_{\uA'}
}.
\labl{II.8} 
} 
In analogy to the quadratically coupled case \ct{Achiman:1985fh, 
GrootNibbelink:1998tz} we have introduced generalizations of the 
connections
\equ{
\gG^\ga_{\gb \gg} \equiv g_\gs^{\;\ga \uga} g_{\gs\, \uga \gb , \gg}, 
 \qquad \gG^A_{B\gg} \equiv g^{A \uA } g_{\uA B, \gg}, \non \\
\labl{II.4} \\
\gG^A_{BC} \equiv g^{A \uA } g_{\uA B, C}, \qquad
\gG^A_{x\gg} \equiv g^{A \uA } g_{\uA x, \gg}
\non
} 
and their conjugates. (There is no object $\gG^A_{xC}$ as a 
similar definition as in eqs.\ \eqref{II.4} just gives 
$\gG^A_{xC} = \gd^A_C$.) Indeed, the metric of the full system 
after this transformation is
\equ{
g_{\ucA' \cA'} = \pmtrx{
g_{\uga'\ga'} & 0
\\
0 & g_{\uA ' A'}
},
\labl{II.9} 
} 
with the effective metric for the $z^\ga, \bar z^{\uga}$ scalars 
given by 
\equ{
g_{\uga \ga} = 
g_{\gs\, \uga \ga} + R_{\bx x \uga\ga}.
\labl{II.10}
} 
In this derivation we have assumed that the metric $g_{\uA A}$ is 
invertable, and we have used the generalized curvature 
$R_{\bx x \uga\ga}$ defined by
\equ{
R_{\bx  x \uga \ga} \equiv 
g_{\bx  x, \uga \ga}
- g_{\bx  B,\uga} g^{B\uB} g_{\uB x, \ga}
= g_{\bx  x, \uga \ga} - \bgG^\uB_{\bx \uga} g_{\uB B}\gG^B_{x\ga}.
\labl{II.10.1}
} 
In the following we also assume that the metric \eqref{II.9} is 
invertable. Notice that the inverse of this transformation is given 
by the same matrices \eqref{II.8} but the primed-indices now are 
downstairs and there  is an additional minus-sign in front of 
the off-diagonal parts.

We could also have chosen the matrices \eqref{II.8} differently 
to block diagonalize the total metric (for example use an 
upper triangle matrix for the first one) but as the metric 
$g_{\gs\uga\ga}$ was already modified (see eq. \eqref{II.7}), 
it is most convenient to include all the other modifications 
in there as well. They can all be combined simply in the curvature 
\eqref{II.10.1}.

Using the connections \eqref{II.4} one can define quite a number 
of generalized curvature components
\equa{
R_{\uga\ga \ugb  \gb } \equiv &  
g_{\gs\, \uga \gg} \lh \gG^\gg_{\ga\gb } \rh_{,\ugb } =
g_{\gs\, \ugb \ga} \lh \bgG^{\ugg}_{\uga \ugb } \rh_{,\gb },
\non \\ 
R_{\uA A \uB B} \equiv &  
g_{\uA  C} \lh \gG^C_{AB} \rh_{,\uB} =
g_{\uC A} \lh \bgG^{\uC}_{\uA \, \uB} \rh_{,B},
\non \\ 
R_{\uA A \uga \ga} \equiv &  
g_{\uA  C} \lh \gG^C_{A\ga} \rh_{,\uga} =
g_{\uC A} \lh \bgG^{\uC}_{\uA \, \uga} \rh_{,\ga},
\labl{II.5} \\
R_{\bx  A \uga\ga} \equiv &
g_{\uC A} \lh \bgG^{\uC}_{\bx \uga} \rh_{,\ga},
\qquad
R_{\uA x\uga\ga} \equiv 
g_{\uA C} \lh \gG^{ C}_{x \ga} \rh_{,\uga},
\non \\
R_{\bx  A \uga B} \equiv &
g_{\uC A} \lh \bgG^{\uC}_{\bx \uga} \rh_{,B},
\qquad
R_{\bA x\uB \ga} \equiv 
g_{\uA C} \lh \gG^{ C}_{x \ga} \rh_{,\uB},
\non}

\nit
Other generalized curvature components either vanish or are 
irrelevant in the following.

Let us mention an important application of the transformation  
diagonalizing the metric to eq.\ \eqref{II.9}.
For several physical applications, like determining whether 
there is soft supersymmetry breaking, one needs to know the 
contracted connection 
\(
\gG_\cA = \gG^\cB_{\cB\cA}
\)
and the Ricci-tensor 
\(
R_{\ucA \cA} = g^{\cB \ucB} R_{\ucB \cB\ucA \cA}
\)
of the full model.
In particular the calculation of the curvature can be very 
tedious even in the setup presented here, and it is hard to 
obtain the Ricci tensor in this way.
However it is well known that the contracted connection and 
the Ricci tensor can be obtained from the determinant $\det g$ 
of the metric 
\equ{
\gG_\cA = (\ln \det g)_{,\cA}, \qquad
R_{\ucA \cA} =(\ln \det g)_{, \ucA \cA}.
\labl{II.5.1}
} 
As the transformation matrices \eqref{II.8} are upper- or 
lower-triangular matrices their determinants are unity.
Therefore we may use the block-diagonal metric \eqref{II.9} 
to calculated the determinant of the full metric:
\(
\det g' = \det g
\). 

There are some further applications of the method discussed in 
the appendix. If the transformations described by the 
matrices \eqref{II.8} are applied to the derivative of the 
coordinates $z^\ga, x^A$ we find that
\eqref{I.4}) 
\equ{
{(\der_\gm Z)'}^{\cA'} =
\pmtrx{
\der_\gm z^{\ga'} \\ \Der_\gm x^{A'}
}
\equiv
\pmtrx{
\der_\gm z^{\ga'} \\ \der_\gm x^{A'} + \gG^{A'}_{x \gb'} 
\der_\gm z^{\gb'}
}.
\labl{II.11.1} 
} 
The derivative $D_\gm x^{A'}$ is covariant under holomorphic 
transformations. From now on we drop the primes on the indices 
if no confusion is possible. Using these definitions the kinetic 
energy of the boson fields $z^\ga$ and $x^A$ can be written as
\equ{
- \cL_{B} = g_{\uga \ga} \der^\gm \bz^{\uga} \der_\gm z^{\ga} 
+ g_{\uA A} \Der^\gm \bx^{\uA} \Der_\gm x^{A}. 
\labl{II.11.2} 
} 
The fermion $\gch^A_L$, the fermionic partner of $x^A$, is 
turned into a covariant vector by the same transformation
\equ{
{\gch'}_L^{A} = \gch_L^A + \gG^A_{x\gb }  \gps^\gb _L \equiv 
 \hat \gch^A_L
\labl{II.12} 
} 
where the hat denotes covariantization.

So far we have only discussed how the metric and covariant 
vectors behave under the transformations described by the matrices 
\eqref{II.8}. This is sufficient to write the kinetic lagrangean 
for the complex scalars in a convenient form. We now turn to the 
calculation of the kinetic lagrangean of the chiral fermions.
These terms in eq.\eqref{O.2.2} involve the covariant derivative 
on the chiral fermions of the full sytem, so we have know the form 
of the covariant derivate on a covariant vector $V^\cA$.
To calculate this we use eq.\eqref{I.8} of the appendix:
\equ{
(\cD_\gm V)_{\ucA'} = {(\cD_\gm V)'}_{\ucA'} 
- \bar U^{\ucE'}_{\ucA' \cE'} g_{\ucE' \cB'} \der_\gm z^{\cE'} V^{\cB'}
+ g_{\ucA' \cE'}  U^{\cE'}_{\cB' \ucE'} \der_\gm \bz^{\ucE'} V^{\cB'}.
\labl{II.14} 
} 
This means we have to calculate the non-vanishing contributions to 
the connection 
\[
\gG^{\prime\cA'}_{\cB'\cC'} = g^{\prime\cA' \ucA'} 
g^\prime_{\ucA' \cB', \cC'}
\]
of the full system 
\equa{
\barr{cc}
{g'}_{\uga'\ga', \cC'} = & \pmtrx{
g_{\uga' \gd'} \gG^{\gd'}_{\ga'\gg'} + R_{\bx  x\uga' \ga'; \gg'} 
+ R_{\bx  B \uga' C} \gG^B_{x\ga'} \gG^C_{x\gg'},
&
R_{\bx x \uga'\ga', C'}
}
\\ & \\ 
{g'}_{\uA 'A', \cC'} = & 
g_{\uA ' B'} \pmtrx{
\hat{\gG}^{B'}_{A' \gg'},
&
\gG^{B'}_{A' C'}
}
\equiv
\pmtrx{
g_{\uA ' B'} \left( \gG^{B'}_{x\gg'} \right)_{, A'}, 
&
g_{\uA ' B'} \gG^{B'}_{A' C'}
}
\earr
\labl{II.15} 
} 
which involves the metric $g_{\uga' \gd'}$ of the transformed 
system. On the r.h.s.\ the index $\cC' = (\gg', C')$ is written out
explicitely using a row-vector notation. The non-vanishing  
components of $U^{\cA'}_{\cB' \ucC'}$ are:
\equ{
U^{A'}_{\ga' \ucC'} =   -g^{A' \uA '}  \pmtrx{
R_{\uA ' x \ugg'\ga'} - R_{\uA ' x \uD' \ga'} 
\bar \gG^{\uD'}_{\bx  \ugg'} 
\\
R_{\uA ' x \uC' \ga'} 
}.
\labl{II.16} 
} 
In these expressions we have made use of a covariant derivative 
in $R_{\bx  x\uga' \ga'; \gg'}$ which is defined in the usual 
way using the connections given in equations \eqref{II.4}, 
whilst we have used the identities
\equa{
\left( \gG^A_{x\gg} \right)_{,C}  = &
 \gG^A_{C\gg} -  \gG^A_{BC} \gG^B_{x\gg}
\equiv 
\hat{\gG}^A_{C \gg},
\non \\
R_{\bx x \uga\ga, B} = &
R_{\bx B \uga\ga} - R_{\bx  C \uga B}\gG^C_{x\gg}.
\labl{II.17}
} 
With this it is easy to give the rewritten covariant derivative 
explicitly. As an application we give here the kinetic terms of 
the supersymmetric lagrangean \eqref{O.2.2} for the chiral 
fermions including covariantizations: 
\equa{
- \cL_{F} = & g_{\uga'\ga'} 
\bgps^{\uga'}_L \lrset\Der\Slashed \gps^{\ga'}_L
+ g_{\uA ' A'} 
\hat{\bgch}^{\uA '}_L \lrset\Der\Slashed \hat \gch^{A'}_L 
\non \\ & \non \\ &
-\left[ 
\left(
R_{\bx x \ugg' \ga', \uA'}
+ 2 R_{\uA ' x \ugg'\ga'} 
- 2 R_{\uA ' x \uD' \ga'} \bgG^{\uD'}_{\bx \ugg'} 
\right) 
\der_\gm \bz^{\ugg'}
\right.
\non \\ & \non \\ &
\qquad \left.
+2 R_{\uA ' x \ubar C' \ga'} 
\Der_\gm \bx ^{\ubar C'} 
\right]
\hat{\bgch}^{\uA '}_L \gg^\gm  \gps_L^{\ga'}
\non \\ & \non \\ & 
+\left[
\left(
R_{\bx x \uga' \gg', A'}
+ 2 R_{\bx A' \uga'\gg'} 
- 2 R_{\bx A' \uga' D'} \gG^{D'}_{x \gg'} 
\right) 
\der_\gm z^{\gg'}
\right.
\non \\ & \non \\ &
\qquad 
\left.
+2 R_{\bx A'\uga' C'} 
\Der_\gm x ^{C'} 
\right]
\bgps_L^{\uga'} \gg^\gm \hat{\gch}_L^{A'}
\non \\ & \non \\ & 
+  \left(
R_{\bx  x\uga' \ga'; \gg'} 
+ R_{\bx  B' \uga' C'} \gG^{B'}_{x\ga'} \gG^{C'}_{x\gg'}
\right) 
\der_\gm z^{\ga'}
\bgps^{\uga'}_L \gg^\gm  \gps_L^{\gg'}
\labl{II.18}
\\ & \non \\ &
-  \left(
R_{\bx  x\uga' \ga'; \ugg'} 
+ R_{\uB' x \uC' \ga'} \bgG^{\uB'}_{\bx\uga'} \bgG^{\uC'}_{\bx\ugg'}
\right) 
\der_\gm \bz^{\uga'}
\bgps^{\ugg'}_L \gg^\gm  \gps_L^{\ga'}
\non
} 
with the covariant derivatives defined in eq.\ \eqref{II.11.1} and
\equ{
\barr{rl}
\Der_\gm \gps^{\ga'}_L \equiv  & \der_\gm \gps^{\ga'}_K
 + \gG^{\ga'}_{\gb '\gg'} \der_\gm z^{\gg'} \gps^{\gb '}_L
\\ & \\ 
\Der_\gm \hat \gch^{A'}_L \equiv & \der_\gm \hat \gch^{A'}_L
  + \hat{\gG}^{A'}_{B'\gg'} \der_\gm z^{\gg'} \hat \gch^{B'}_L
  + \gG^{A'}_{B'C'} \der_\gm x^{C'} \hat \gch^{B'}_L.
\earr
\labl{II.19} 
} 
The four-fermion terms can be calculated by using eq.\ \eqref{I.15} 
in the appendix. 

\section{Geometry of matter in Grassmannian Models}
\label{VIII} 

The previous section was devoted to the question of how one could 
make the metric of the combined system of matter and $\gs$-model 
fields block diagonal. In doing so we noticed, that these 
techniques can be applied to make all kinds of functions of 
fields covariant under the isometries of the underlying 
$\gs$-model. In this section we show how these methods may work 
in practice with the example discussed in section \ref{VII} of 
consistent Grassmann $\gs$-models with the field content of the 
standard model with one generation. Our starting point is the 
quadratically coupled matter \Kh\ potential \eqref{VI.11}.
Using the results of section \ref{II} we have computed the 
connections \eqref{II.4}
\equa{
\gG_{(ia)(jb)}^{(kc)}
= & -\get f^2 
\lh \gd^c_b \gd^k_i (\tg \bQ)_{aj} + \gd^c_a \gd^k_j (\tg \bQ)_{bi} \rh, 
\non\\ & \non\\
\gG_{E(jb)} = & -e \get f^2 (\bQ g)_{jb} E,
\non\\ & \non\\
\gG^k_{L(jb)} = & - \get f^2 
\lh l (\bQ g)_{bj} L^k + \gd^k_j  (\bQ g L)_{b} \rh, 
\labl{VIII.3}
\\ & \non\\
\gG^c_{D(jb)} = & - \get f^2 
\lh d (\tg \bQ)_{bj} D^c + \gd^c_b  (D\tg \bQ)_{j} \rh. 
\non
} 
The connection for $U$ is similar to the one for $D$, and the 
connections for the Higgses $H^\pm$ are similar to the one for 
$L$. (In models with more generations, the quark doublets $Q'$ 
have the same connection as the $\gs$-model field $Q$.)
To make a distinction between indices referring the original 
$\gs$-model fields $Q^{ia}$ and matter indices $a$ and $i$, 
we write $(ai)$ for the former ones. Notice that the normal gauge, 
in which all connections vanish, coincides with the unitary gauge 
$Q = 0$. Because of the global $U_\get(2,3)$ invariance, the vacuum 
can always be studied in the normal gauge by setting $<Q> = 0$.
Using these connections, one obtains the covariant chiral 
fermions by eq. \eqref{II.12}, for example
\equa{
\hat e^{c}_{L}  \equiv e^{c\prime}_{L}  
= &e^c_L - e \get f^2 E\ \tr_m \bQ g q^{\;}_L,
\non \\ &  \labl{VIII.3.1} \\ 
\hat l_L^{i} \equiv l_L^{i\prime} 
= & l_L^i - \get f^2 \left(   
l L^i\ \tr_m \bQ g q^{\;}_L + (q_L \bQ g L)^i
\right).
\non
}
Because we only consider quadratically coupled matter here, 
we have $\gG^A_{x\ga} = x^B \gG^A_{B\ga}$ and $\gG^A_{BC} = 0$.
For the same reason most of the curvatures of eq.\eqref{II.5} 
are related; we find
\equa{
R^{(ia)\;\;\; (kc)}_{\;\;\;\; (bj)\;\;\; (dl)} = &
- \get f^2 \lh
g^{\;(ic)}_{\gs \;\;\; (bj)}\,  g^{\;(ka)}_{\gs \;\;\; (dl)} 
+
g^{\;(ka)}_{\gs \;\;\;(bj)} \, g^{\;(ic)}_{\gs \;\;\; (dl)} 
\rh,
\non\\ & \non\\
R_{\bE E \;\;\;\; (ia)}^{\;\;\;\;\; (bj)} = &
- \get f^2  
e K_E g^j_{\; i} \tg^{\; b}_a\,
= -\get e f^2 K_E g_{\gs\;\;\;\; (ia)}^{\; (bj)}
\non\\ & \non\\
R_{\bL L \;\;\;\; (ia)}^{\;\;\;\; (bj)} = &
- \get f^2 \lh 
l K_L  g^j_{\; i}   + ([L\bL]\,g)_i^j
\rh \tg^{\; b}_a,
\labl{VIII.4}
\\ & \non\\
R_{\bD D \;\;\;\; (ia)}^{\;\;\;\;\; (bj)} = &
- \get f^2 \lh 
d K_D  \tg^{\; b}_a  + (\tg\, [\bD D])^b_a 
\rh  g^j_{\; i},
\non
} 
using the notation $[L\bL]$, etc., of section \ref{VI}.
The metric $ G_{\gs\;\;\;(ia)}^{(bj)}$ of the $\gs$-model fields 
$Q$ and $\bQ$  in the presence  of matter multiplets $E, L, D, U$ 
becomes
\equa{
G_{\gs\;\;\;\; (ia)}^{\;(bj)} \equiv \ & 
g_{\gs\;\;\;\; (ia)}^{\; (bj)}
+ 
\sum_x R_{\bx x \;\;\;\;(ia)}^{\;\;\;\; (bj)} 
 = 
\ga\left( 
g \otimes \tg +  g A \otimes \tg + g \otimes B \tg 
\right)_{\;\;\;\; (ia)}^{(bj)},
\labl{VIII.5}
} 
using eq.\eqref{II.10} as well as the curvatures \eqref{VIII.4} 
with the short-hand notation 
\equa{
\ga = & 1 - \get f^2 \sum_x q_x K_x,
\non \\
A = & - \get f^2 \ga\inv 
\lh [L \bL] + [H^+ \bH^+] + [H^- \bH^-]  \rh, 
\labl{VIII.5.1} \\ 
B = & - \get f^2 \ga\inv 
\lh [\bD D] +  [\bU U] \rh.
\non
} 
Notice that in the unitary gauge $Q = 0$ the metric $G_\gs$ does 
not reduce to the metric without matter coupling $g_\gs$ evaluated 
at $Q = 0$. The inverse of this metric can be written as infinite 
sum of tensor products
\equ{
G\inv_\gs = \ga\inv \sum_{n = 0}^\infty 
(1 + A)^{-n-1} A^n g\inv  \otimes \tg\inv B^n (1 + B)^{-n-1}.
\labl{VIII.2}
} 
It turns out that it is very profitable to express other 
quantities using the covariant objects defined in section \ref{II} 
as well.
First of all we find that the first order derivative of the \Kh\ 
potential simplifies to
\equ{
K_{,\cA'} = \left(
(\tg \bQ)_{(ai)},  
\bE g^{(E)},
(\bL g^{(L)})_i,  
(\bH^+ g^{(H^\pm)})_i,  
(\tg^{(D)} \bD)_a, 
(\tg^{(U)} \bU)_a
\right).
\labl{VIII.6}
} 
The full metric in the transformed system is given by
\equ{
G' = \mbox{diag}\left(
G_\gs, g^{(E)}, g^{(L)}, g^{(H^+)}, g^{(H^-)}, \tg^{(D)}, \tg^{(U)}
\right),
\labl{VIII.7}
} 
where $G_\gs$ is given by \eqref{VIII.5}.

\section{Vacua of the Grassmannian standard model}
\label{IX}

Section \ref{VI.sub1} discussed a chiral anomaly-free Grassmannian 
model with the fermion particle spectrum of the standard model.
We now discuss the possible vacuum configurations of this 
model. Grassmannian models with doubling have been studied 
in a supergravity background \cite{Goto:1992hh}, but the authors 
did not include superpotential terms which can alter their 
claim that the fermion masses are of the order of the 
gravitino mass. Using the geometrical results of section 
\ref{VIII} we can discuss the vacuum solutions of this 
model in an elegant and straightforward fashion.

Before going into the details of the model we first observe 
that --barring non-trivial topological effects \ct{sjwinpre}-- 
the vacuum can always be chosen such that 
${\langle Q \rangle = 0}$.
As the vacuum expectation values 
of $Q$ and $\bQ$ are constants, they can be set to zero by a 
global gauge 
transformation. Notice that $\langle Q \rangle = 0$ 
is indeed a vacuum solution, because in the scalar potential $Q$ and 
$\bQ$ always appear together.

In the supergravity background the consistent model of 
$U_\get(2,3)/U(2)\times U(3)$ with the chiral fermion content 
of the standard model should satisfy at least the following 
requirements in order not contradict the standard model 
phenomenology: the gauge group $SU(3)\times U_{em}(1)$ is 
unbroken, and the gauginos and the complex scalar bosons 
should acquire masses above the scale of the gauge bosons 
and the chiral fermions.

Here we analyze the restrictions resulting from the 
electroweak symmetry breaking. The subgroup $SU(3) \times 
U_Y(1) \times SU(2)$ is gauged and the generator 
$Q_{em} = \frac 12 Y_w + I_3$ is unbroken. Therefore all 
singlets under $SU(2)$ should vanish in the vacuum; this 
holds in particular for the covariant superpotential $\cW$. 
Furthermore we demand that $B-L$ is a good symmetry, also below 
the electroweak symmetry breaking scale. 
Only neutral parts of the Higgs $SU(2)$ doublets may acquire a 
vacuum expectation value
\equ{
<H^+> = \pmtrx{0 \\ H^+_0},
\qquad 
<H^-> = \pmtrx{H^-_0 \\ 0}.
\labl{IX.1}
} 
The Killing potentials of the $Y$-charge and the weak-isospin
\equa{
\cM_Y = & - \frac 6{\get f^2} 
+ 15\lh |H_0^+|^2 - |H_0^-|^2 \rh,
\non \\ & 
\labl{IX.2} \\ 
\cM_{I_3} = & \half \lh |H_0^-|^2 - |H_0^+|^2 \rh,
\non
} 
are the only Killing potentials which do not necessarily vanish. 
The non-vanishing part of the scalar potential due to the 
$D$-terms is given by
\equ{
V_D = \frac 12 g_Y^2 \cM_Y^2 + \frac 12 g_2^2 \cM_{I_3}^2.
\labl{IX.2.1}
}
\nit
When only the standard model gauge group is gauged, the gauge couplings
are independend. We denote the $U(1)$ gauge coupling constant by $g_Y$, 
the gauge coupling constants for $SU(2)$ and $SU(3)$ by $g_2$, $g_3$. 
We observe, that there always is a $D$-term supersymmetry and internal 
symmetry breaking, and the minimum of the potential occurs at
\equ{
|H_0^+|^2 - |H_0^-|^2 = - 2 \cM_{I_3} = 
\frac{15g_Y^2 }{(15 g_Y)^2 + \frac 14 g_2^2} 
\lh
 \frac 6{\get f^2}
\rh
\labl{IX.3.1}
} 
in both cases ($\get = \pm 1$). The other Killing potential 
takes the value
\equ{
\cM_{Y} = 
\frac{ - \frac 14 g_2^2 }{(15 g_Y)^2 + \frac 14 g_2^2} 
\lh
\frac 6{\get f^2}
\rh.
\labl{IX.3.2}
} 

In section \ref{XI} we discussed special requirements which 
have to be fulfilled in order for $<\cW> = 0$ to be allowed.
We investigate the consequences of these conditions.
We assume that the covariant superpotential can be written as
\(
\cW = w f^3 W,
\)
eq.\ \eqref{O.31}. (In general this may be a sum of such 
products.) For the invariant superpotential $W$  one can take 
\eqref{VI.23}. For the compensating superpotential $w$  we can 
choose between two compensating superpotentials, see 
\eqref{VI.22} where also their charges can be found. Because of 
the strong restriction that the covariant superpotential has to 
vanish in the vacuum, it follows that the gravitino 
mass vanishes. Therefore soft supersymmetry breaking masses 
can only arise due to the non-linear nature of the model.
The minimal condition for which the 
covariant superpotential may vanish is that
\equ{
0 < - \frac 1\go = - \frac {\gk^2}{q \get f^2}.
\labl{IX.4}
} 
This requirement specifies which version of $U_\get(2,3)$ one  
should use. The compensating superpotential $w_E$ is relevant in 
the non-compact version ($\get = -1$) 
as the charge $q_E$ is positive, to incorporate 
proper electroweak symmetry breaking. 
However the charge of 
$w_{L-}$ is negative, so it should be used in the compact 
version ($\get = 1$). 
When $0 < -\frac 1\go \leq 1$ the derivatives 
of the covariant 
superpotential should vanish as well.

We will now analyse the different cases in more detail, 
starting with $w_E$. 
In the case that 
\(
-\frac 1\go > 1
\)
the scalar potential is always at its minimum if eq.\ \eqref{IX.3.1} 
is satisfied, but $\tan \gb$ is arbitrary. 
When $0 < -\frac 1\go \leq 1$ we find 
that in addition
\equ{
 \ga - \gm |H_0^+||H_0^-| = 0.
\labl{IX.5}
} 
From this equation together with \eqref{IX.3.1} we get a 
prediction for the ratio of the two VEVs of the Higgses
\equ{
\tan^2 \gb \equiv \frac{|H^+_0|^2}{|H^-_0|^2} = 
\frac{
\sqrt{ \cM_{I_3}^2 +  \lh \ga/\gm \rh^2} -  \cM_{I_3}
}{
\sqrt{ \cM_{I_3}^2 +  \lh \ga/\gm \rh^2} +  \cM_{I_3}
},
\labl{IX.6}
} 
where $\cM_{I_3}$ is given by eq. \eqref{IX.3.1}.

Finally we consider the case of $w = w_{L-}$.
When $ 1 < -\frac 1\go$, 
the only restriction on $H_0^\pm$ is eq.\ \eqref{IX.3.1}; 
$\tan \gb$ remains undetermined. 
However when $0 <-\frac 1\go \leq 1$, the vanishing of the derivatives 
of the covariant superpotential demands that either 
$H_0^- = 0$ or eq. \eqref{IX.5} is satisfied. 
There are two 
inequivalent vacua which both break the electroweak symmetry. 
First of all
\equ{
H_0^- = 0, \quad
|H_0^+|^2 = - 2 \cM_{I_3} = 
\frac{15g_Y^2 }{(15 g_Y)^2 + \frac 14 g_2^2} 
\lh
 \frac 6{\get f^2}
\rh, 
\labl{IX.8}
} 
which gives the unacceptable result $\tan \gb = \infty$.
The other vacuum solution leads to a $\tan \gb$ as given in 
eq.\ \eqref{IX.6}.

\section{Conclusions}
\label{X} 

Effective field theory may serve as a powerful tool in the 
study of physics beyond the standard model up to the intrinsic 
cut-off scale, which could be as large as the Planck scale.
If the theory is realized in a broken phase, the symmetries 
are non-linear. For $N=1$ supersymmetric theories this involves 
the study of \Kh\ manifolds. \Kh ian coset models provide a 
class of interesting examples, but unfortunately in their 
simplest version these models are inconsistent. Until recently 
the only known method to make these models consistent in a 
supersymmetric way, was by doubling the spectrum by adding 
mirror chiral superfields. The phenomenology of these doubling 
models is unsatisfactory as the fermions can easily get 
masses of the cut-off scale. If a renormalizable supersymmetric 
field theory is plagued by anomalies, one adds extra matter 
representations with the appropriate quantum numbers.  
When matter is coupled to \Kh\ models, this can be done 
similarly if the charges of the matter superfields can be 
manipulated freely. In ref.\ \cite{GrootNibbelink:1998tz} 
we showed that it was possible to do this, and construct 
consistent supersymmetric $\gs$-models without resorting 
to mirror chiral superfields. The crucial step is that one 
can couple a singlet to the \Kh\ manifold using the holomorphic 
functions $\cF_i$ in which the \Kh\ potential transforms. 
Once it was understood how to couple a singlet with an 
arbitrary $U(1)$ charge to the $\gs$-model, the door was 
open to change the charges of other matter representations 
as well using rescaling of these matter fields by a non-trivial 
singlet. 

In this article we have reviewed and extended these ideas. 
The Killing potentials were used also to give a non-trivial
example of a non-standard, non-minimal gauge kinetic function. 
We have discussed in detail the coupling of \Kh\ models with 
additional matter to supergravity. We showed that in 
supergravity the rescaling of the matter fields is a 
consequence of their Weyl-weights and the covariance of the 
\Kh\ potential. The compensating singlet of superconformal 
gravity can be used to cancel the transformation of the \Kh\ 
potential, before reducing it to supergravity. This singlet 
is used to set the Weyl-weights of the matter fields to zero.
Doing this introduces the same additional transformation 
rules for the matter fields. Because of the transformation 
properties of the compensating singlet in supergravity, the 
superpotential has to be covariant as well. Using this covariant 
superpotential one can construct an invariant \Kh\ potential. 
With the auxiliary fields of the gauge multiplets coupling to 
the scalars via the Killing potentials we obtain additional 
contributions to the scalar potential.

The study of the vacuum configurations of these models implies 
that either the \Kh\ $U(1)$ isometry is broken or additional 
requirements have to be satisfied. Either there is a relation 
between the cut-off scale $f\inv$, the Planck scale and the 
transformation properties of the covariant superpotential or 
there are more requirements on the VEVs of the scalar fields.
Another consequence is that the gravitino mass vanishes in 
the case of an unbroken \Kh\ $U(1)$.

The consistent system of $\gs$-model and matter superfields 
can become quite complicated. In particular the various 
irreducible representations have mixed kinetic terms 
as the metric is not block diagonal. By applying a 
non-holomorphic transformation on covariant objects, 
like chiral fermions of the full model and derivatives of the 
scalars, it is possible to block-diagonalize the metric. This 
transformation also turns non-covariant objects, like the 
chiral fermions belonging to the matter sectors, into covariant 
ones. This method is explained in section \ref{II} and the 
geometrical background can be found in the appendix. 
The automatic covarantizations are convenient consequences 
of this method, but other calculations are simplified as well.

All these different aspects are illustrated by the example of 
Grassmannian coset models $U_\get(m,n)/U(m)\times U(n)$.
The properties of matter coupling to Grassmannian \Kh\ 
manifolds are described by using left- and right-covariant 
real composite superfields. This offers many different ways 
to construct non-equivalent invariant \Kh\ potentials for 
the matter fields. The algebra of isometries of the $\gs$-models 
is discussed in detail, identifying the \Kh\ $U(1)$ charge and 
a central charge. 

At the classical level the isometries can be gauged by a 
straightforward procedure. Non-standard non-minimal kinetic terms 
for the gauge fields were constructed, but supersymmetry requires 
them to be accompanied by higher-deriv\-ative terms involving the 
components of the physical scalar supermultiplets. Some of these 
terms disappear in the broken phase of local gauge symmetries,
in which case the $U(1)$-coupling constant is renormalized 
w.r.t.\ the remaining part of the gauge group. 

As a practical illustration of the cancellation of the anomalies 
in a Grassmannian coset models, we discussed a model of the standard 
model where the superpartner of the quark-doublet is interpreted as 
the coordinates of the coset $U_\get(2,3)/U(2)\times U (3)$. We 
showed how on this \Kh\ manifold matter representations could be 
added in such a way that the chiral fermion sector of the model 
coincides with the standard model. As the covariant superpotential 
plays an important role in supergravity, the construction 
of the superpotential was discussed in some detail. The power 
of the non-holomorphic transformation on covariant objects 
was illustrated for the calculation of the metric and 
first derivative of the \Kh\ potential; we obtained the expressions 
for the covariant chiral matter fermions in this way. We showed 
that in supergravity the Grassmannian version of the standard model 
leads to phenomenologically acceptable results only in very specific 
cases. The compensating superpotential is either $w_{L-}$ or 
$w_{E}$, see eqs. \eqref{VI.22}. 
The former leads to 
two inequivalent vacua, of which only one is acceptable, as 
in the other case one of the two Higgses has a vanishing VEV. 
The covariant superpotential $w_E$ also gives rise to an 
acceptable vacuum, but with a different value for $\tan \gb$.
\\
\\
{\large\bf Acknowledgment}
\\
\\
We would like to thank B.J.W. van Tent for reading 
the manuscript and for helpful comments.

\appendix

\section{Appendix: Geometry of \Kh\ manifolds}
\label{I}
In section \ref{O} we review the structure of $N=1$ supersymmetric 
models in 4 dimensions and come across various geometrical objects 
like the metric, connection and curvature. These geometrical objects 
are used there as convenient short-hand to write the lagrangean in a 
compact form. They have very specific functions in the supersymmetric 
lagrangean: the kinetic energy of the scalars and the fermions are 
described by the metric. The Dirac operator that acts on the chiral 
fermions involved the connection and the four-fermion interactions 
couple via the curvature after the auxiliary fields are removed, 
see eq.\eqref{O.2.2}. Section \ref{II} discusses a few of these 
applications like how to define fermions in chiral matter multiplets 
in order that they transform covariantly and how to make the metric 
block diagonal. In this appendix we look at these objects 
from a geometrical point of view but keeping physical applications 
in mind. We consider a \Kh\ manifold described (locally) by a 
\Kh\ potential $\cK(\bZ, Z)$ and treat the superpartners 
$\gps_L^\cA$ of $Z^\cA$ in exactly the same way as covariant 
fields that live on this \Kh\ manifold. Various transformations 
that can act on covariant fields are studied in this appendix. 
In section \ref{II} these transformations are used to cast the 
supersymmetric lagrangean involving matter multiplets into a 
form depending only on physical covariant fields. 

Since a \Kh\ manifold is complex, the coordinate transformations 
preserving the complex structure are holomorphic 
\equ{
Z^\cA \lra {Z'}^{\cA'}  = \cR^{\cA'}(Z), \qquad 
\bZ^{\ucA} \lra {\bZ}^{\prime\ucA'}  = \bcR^{\ucA'}(\bZ).
\labl{I.1}}

\nit
Any object $V^\cA$ (and its conjugate $\bar V^{\ucA}$) 
transforming as 
\equ{
V^\cA \lra {V'}^{\cA'}  = X^{\cA'}_{\;\cA} (Z) V^\cA , \qquad 
\bV^{\ucA} \lra {\bV}^{\prime \ucA'}  = \bX^{\ucA'}_{\;\ucA} 
(\bZ) \bV^{\ucA}
\labl{I.2} 
} 
under the holomorphic coordinate transformations with 
\equ{
X^{\cA'}_{\;\cA} (Z)  =  \cR^{\cA'}_{\;,\cA} (Z), \quad
  \bX^{\ucA'}_{\;\ucA} (Z) =  \bcR^{\ucA'}_{\;,\ucA} (Z) 
\labl{I.3}
}
is called a covariant vector of the \Kh\ manifold. 
In the context of supersymmetric $\gs$-models many covariant 
vectors are encountered, to name a few: 
the derivatives $\der_\gm Z^\cA$, the differentials $d Z^\cA$ 
and the superpartners $\gps_L^\cA$ of $Z^\cA$.

The coordinate transformations \eqref{I.1} generally do not 
leave the metric of the \Kh\ manifold invariant, only the 
$S$-matrix of the field theory described by these coordinates.
The coordinate transformations that do leave the metric 
invariant are called isometries. 

On the covariant vectors the transformation rules 
\eqref{I.2} we can consider more general transformations
\equ{
V^\cA \lra {V'}^{\cA'}  = X^{\cA'}_{\;\cA} (\bZ, Z) V^\cA,
\labl{I.4}}

\nit
where $X^{\cA'}_{\;\cA}(\bZ, Z)$ are possibly non-holomorphic 
functions. This type of transformations can be used to make 
the physical content of a field theory more transparent, as 
is illustrated in section \ref{II}. The first thing to note 
is that these transformations can not be generated by 
non-holomorphic coordinates transformations because they 
would introduce terms involving $\bV^\ucA$ in eq.\eqref{I.4} 
too. Therefore the transformations \eqref{I.4} can only be defined 
on the level of covariant vectors and geometrical objects like 
the metric: ${V'}^{\cA'}$ is nothing but a short-hand for the 
expression $ X^{\cA'}_{\;\cA} (\bZ, Z) V^\cA$ for the covariant 
vector $V^\cA$. 
In the following we study how the transformations \eqref{I.4} 
change the appearance of formulae involving the metric, 
connection and curvature.

If we demand that the metric defines an invariant inner 
product for covariant vectors, it must transform as
\equ{
g_{\ucA \cA} \lra  {g'}_{\ucA' \cA'} = \bX_{\ucA'}^{\;\ucA} 
X_{\cA'}^{\;\cA} g_{\ucA \cA}
\labl{I.5}} 
where $X^{\;\cA}_{\cA'}(\bZ, Z)$ is the inverse of 
$X^{\cA'}_{\;\cA} (\bZ, Z)$.

A word about our notation is in order here: let $A_\cA$ 
be any object with one index down, not necessarily a vector; 
it may be a function of covariant vectors and derivatives. 
Applying \eqref{I.4} to all covariant 
vectors transforms $A_\cA$ into ${A'}_{\cA'}$. 
One can also just contract $A_\cA$ with the matrix 
$X_{\cA'}^{\;\cA}$ this is denoted by 
\(
A_{\cA'} = X_{\cA'}^{\;\cA} A_\cA.
\) 
In the case of covariant vectors and the metric 
${g'}_{\ucA' \cA'} = g_{\ucA' \cA'}$ these two definitions 
coincide but this is not true in general.
(When there is no confusion possible, like with covariant 
vectors or the metric, we drop the prime on the symbol itself.)

\nit
The prime example where this is not the case is the connection

\equ{
\barr{ccc}
{\gG}^{\cA}_{\cB\cC} \lra {\gG'}^{\cA'}_{\cB'\cC'} & = &  
{\gG}^{\cA'}_{\cB'\cC'} + {U}^{\cA'}_{\cB'\cC'} 
+ g^{\cA' \ucB'}  \bar U^{\ucA'}_{\ucB' \cC'} g_{\ucA' \cB'}  
\\ & & \\
{\bgG}^{\ucA}_{\ucB \,\ucC} \lra {\bgG}^{\prime \ucA'}_{\ucB'\ucC'} & 
 = & {\bgG}^{\ucA'}_{\ucB'\ucC'} + {\bU}^{\ucA'}_{\ucB'\ucC'} 
+ g^{\cB' \ucA'}  U^{\cA'}_{\cB' \ucC'} g_{\ucB' \cA'}
\earr
\labl{I.6} 
} 
with 
\equ{
\barr{c}
{U}^{\cA'}_{\cB'\cC'} = X_\cA^{\cA'} X^\cA_{\cB', \cC} X^\cC_{\cC'}, 
\qquad
\bar{U}^{\ucA'}_{\ucB'\ucC'} = 
\bar X_{\ucA}^{\ucA'} X^{\ucA}_{\ucB',\ucC} 
X^{\ucC}_{\ucC'},
\\  \\
U^{ \cA'}_{\cB' \ucC'} = X^{\cA'}_{\cA}
X^{\cA}_{ \cB', \ucC} X^{\ucC}_{\ucC'},
\qquad
\bar U^{\ucA'}_{\ucB' \cC'} = \bar X^{\ucA'}_{\ucA}
\bar X^{\ucA}_{\ucB', \cC} X^\cC_{\cC'}.
\earr
\labl{I.7} 
} 
Notice that the third term in equations \eqref{I.6} vanishes if 
the transformations are holomorphic. Here we see clearly that 
the connection is not a tensor even in the case of holomorphic 
transformations. But this exactly enables us to define a covariant 
derivative $\cD_\gm$ for covariant vectors 
\(
\cD_\gm V^\cA \equiv \der_\gm  V^\cA+ \gG^\cA_{\cC\cB} 
\der_\gm Z^\cB V^\cC.
\)
However it is only covariant under holomorphic transformations 
but not under eq.\eqref{I.4}; indeed
\equ{
{(\cD_\gm V)'}^{\cA'} = \cD_\gm V^{\cA'} 
+ g^{\cA' \ucB'}  \bU^{\ucA'}_{\ucB' \cB'} g_{\ucA' \cC'} 
\der_\gm Z^{\cB'} V^{\cC'}
-  U^{\cA'}_{\cC' \ucB'} \der_\gm \bZ^{\ucB'} V^{\cC'}.
\labl{I.8} 
} 
The second term on the r.h.s. follows from eq.\eqref{I.6} and 
the third compensates for the fact that the ordinary derivative 
$\der_\gm$ within $\cD_\gm$ can hit the transformation matrix 
$X_\cA^{\cA'}$ which may also depend on $\bZ^{\ucA}$.
The first term on the r.h.s. is of the same form as what one 
would get if the transformations \eqref{I.4} are holomorphic.
The last two terms involve $U$ and $\bU$'s with mixed indices 
indicating the non-holomorphic nature of \eqref{I.4}.

Finally we investigate how the transformations \eqref{I.4} 
influence the curvature. The calculation follows the same 
line as above, but now it is really convenient to separate 
terms which do not have mixed transformations involving $U$ 
and $\bU$. With this separation one can identify which terms 
behave as if the transformations \eqref{I.4} are holomorphic. 
We call these terms holomorphic and indicate them with a 
superscript $H$. The remaining terms have $U$'s and $\bU$'s 
with mixed indices. They are called non-holomorphic and are 
indicated by a superscript $N$.

As the curvature is a tensor under holomorphic transformations, 
the holomorphic part $R^H$ also transforms as a tensor under 
\eqref{I.4}. By identifying the holomorphic and non-holomorphic 
parts we find
\equ{
\barr{cc}
{R'}_{\ucA'\cA' \ucB' \cB'} = & {R^H}_{\ucA'\cA' \ucB' \cB'} 
+ {g^N}_{\ucA'\cA', \ucB' \cB'}
- {g^H}_{\ucA' \cC', \cB'} g^{\cC' \ucC'} {g^N}_{\ucC' \cA', \cB'} +
\\ & \\
&- {g^N}_{\ucA' \cC', \cB'} g^{\cC' \ucC'} {g^H}_{\ucC' \cA', \cB'} 
+ {g^N}_{\ucA' \cC', \cB'} g^{\cC' \ucC'} {g^N}_{\ucC' \cA', \cB'}. 
\earr
\labl{I.9} 
} 
As we already know how the holomorphic part of the curvature 
transforms, we only have to consider the terms with non-holomorphic 
transformations. In these terms replace the remaining holomorphic 
like parts ${g^H}_{\ucA' \cC', \cB'}$ by $(g' - g^N)_{\ucA' \cC', \cB'}$. 
In section \ref{II} we are not so much interested in the transformed 
curvature itself, but more in having a simple way to write expressions 
involving the curvature, like the four-fermion terms. Therefore we 
write
\equ{
{R'}_{\ucA'\cA' \ucB' \cB'} = {\hat R'}_{\ucA'\cA' \ucB' \cB'} 
+ {g'}_{\ucA'\cA', \ucC'} \bU^{\ucC'}_{\ucB'\cB'}
\labl{I.14}}
 
\nit
and notice that the second term depend on the order of the indices 
$\ucB'$ and $\cB'$ and where the first is given by
\equa{
{\hat R'}_{\ucA'\cA' \ucB' \cB'} = &
{R}_{\ucA'\cA' \ucB' \cB'} 
+ g_{\ucD' \cA'} \bar W^{\ucD'}_{\ucA'\ucB'\cB'}
+ g_{\ucA' \cD'} W^{\cD'}_{\cA'\ucB'\cB'}
\non \\ & \non \\ &
+ g_{\ucA' \cD'} U^{\cD'}_{\cC'\ucB'} g^{\cC'\ucC'}
 \bar U^{\ucD'}_{\ucC'\cB'}  g_{\ucD' \cA'}
- \bar U^{\ucD'}_{\ucA'\cB'}  g_{\ucD' \cD'} U^{\cD'}_{\cA'\ucB'}
\labl{I.15} \\ & \non \\ &
- g_{\ucA' \cD'} \left(
- {\gG'}^{\cD'}_{\cE' \cB'} 
U^{\cE'}_{\cA'\ucB'}
+
{U}^{\cD'}_{\cE' \ucB'} 
{\gG'}^{\cE'}_{\cA' \cB'}
+
{U}^{\cD'}_{\cE' \cB'} 
U^{\cE'}_{\cA'\ucB'}
\right)
\non \\ & \non \\ & 
- g_{\ucD' \cA'} \left(
-\bar {\gG'}^{\ucD'}_{\ucE' \ucB'} 
\bar U^{\ucE'}_{\ucA'\cB'} 
+ 
{\bar U}^{\ucD'}_{\ucE'  \cB'} 
\bar {\gG'}^{\ucE'}_{\ucA'\ucB'} 
+
{\bar U}^{\ucD'}_{\ucE' \ucB'} 
\bar U^{\ucE'}_{\ucA'\cB'} 
\right).
\non
}

\nit
Here $W$ is defined as 

\equ{
W^{\cD'}_{\cA'\ucB'\cB'} = 
X^{\cD'}_{\cD} X^{\cD}_{\cA', \ucB\cB} \,
\bX^{\ucB}_{\ucB'}X^{\cB}_{\cB'}
\labl{I.16}}

\nit
and similarly for $\bar W$.


\providecommand{\href}[2]{#2}\begingroup\raggedright
\endgroup

\end{document}